\documentstyle[11pt,fleqn]{article}

\def\citen#1{%
\edef\@tempa{\@ignspaftercomma,#1, \@end, }
\edef\@tempa{\expandafter\@ignendcommas\@tempa\@end}%
\if@filesw \immediate \write \@auxout {\string \citation {\@tempa}}\fi
\@tempcntb\m@ne \let\@h@ld\relax \let\@citea\@empty
\@for \@citeb:=\@tempa\do {\@cmpresscites}%
\@h@ld}
\def\@ignspaftercomma#1, {\ifx\@end#1\@empty\else
   #1,\expandafter\@ignspaftercomma\fi}
\def\@ignendcommas,#1,\@end{#1}
\def\@cmpresscites{%
 \expandafter\let \expandafter\@B@citeB \csname b@\@citeb \endcsname
 \ifx\@B@citeB\relax 
    \@h@ld\@citea\@tempcntb\m@ne{\bf ?}%
    \@warning {Citation `\@citeb ' on page \thepage \space undefined}%
 \else
    \@tempcnta\@tempcntb \advance\@tempcnta\@ne
    \setbox\z@\hbox\bgroup 
    \ifnum\z@<0\@B@citeB \relax
       \egroup \@tempcntb\@B@citeB \relax
       \else \egroup \@tempcntb\m@ne \fi
    \ifnum\@tempcnta=\@tempcntb 
       \ifx\@h@ld\relax 
          \edef \@h@ld{\@citea\@B@citeB}%
       \else 
          \edef\@h@ld{\hbox{--}\penalty\@highpenalty \@B@citeB}%
       \fi
    \else   
       \@h@ld \@citea \@B@citeB \let\@h@ld\relax
 \fi\fi%
 \let\@citea\@citepunct
}
\def\@citepunct{,\penalty\@highpenalty\hskip.13em plus.1em minus.1em}%
\def\@citex[#1]#2{\@cite{\citen{#2}}{#1}}%
\def\@cite#1{$^{\:#1}$}
\def\thebibliography#1{\baselineskip 1pt\section*{ }
  \addcontentsline{toc}{section}{Note}

\list{\arabic{enumi}.}{\settowidth\labelwidth{[#1]}\leftmargin\labelwidth
    \advance\leftmargin\labelsep
    \usecounter{enumi}}
    \def\newblock{\hskip .11em plus .33em minus -.07em}
    \sloppy\clubpenalty4000\widowpenalty4000
    \sfcode`\.=1000\relax}

\newcommand{\preprint}[1]{\hfill{\sl preprint - #1}\par\bigskip\par\rm}
\def\titolo{\par\bigskip\begin{center}\bf\LARGE}
\def\endtitolo{\end{center}\par\bigskip\par\rm\normalsize}
\def\instit{\begin{center}\large}
\def\endinstit{\end{center}\rm\normalsize}
\newcommand{\dip}{\smallskip Dipartimento di Fisica,
Universit\`a di Trento}
\newcommand{\infn}{\smallskip Istituto Nazionale di Fisica Nucleare,\\
Gruppo Collegato di Trento,\\ 38050 Povo (TN)
Italia}
\newcommand{\dinfn}{\dip\\ and \infn}
\newcommand{\btit}{\begin{titolo}}
\newcommand{\etit}{\end{titolo}}

\newcommand{\Idinfn}{\begin{instit}\dinfn\end{instit}}

\renewcommand{\author}[1]{\begin{center}\Large #1\end{center}}
\renewcommand{\date}[1]{\par\bigskip\par\sl\hfill #1\par\medskip\par\rm}
\newcommand{\pacs}[1]{\smallskip\noindent{\sl PACS number(s):
\hspace{0.3cm}#1}\par\bigskip\rm}
\newcommand{\babs}{\par\begin{description}\item{~}\it}
\newcommand{\eabs}{\par\end{description}\par\medskip\rm}

\renewcommand{\vec}[1]{{\bf #1}}       
\newcommand{\nn}{\nonumber}            
\newcommand{\beq}{\begin{eqnarray}}    
\newcommand{\eeq}{\end{eqnarray}}      
\newcommand{\beqn}{\begin{eqnarray}}   
\newcommand{\eeqn}{\end{eqnarray}}     
\newcommand{\R}{\mbox{$I\!\!R$}}                 
\newcommand{\Z}{\mbox{$Z\!\!\!Z$}}               
\newcommand{\C}{\mbox{$I\!\!\!\!C$}}             
\newcommand{\lap}{\nabla}                        
\newcommand{\lrup}[1]{{\stackrel{\leftrightarrow}{#1}}}
\newcommand{\al}{\alpha}
\newcommand{\be}{\beta}

\newcommand{\ep}{\varepsilon}

\begin{document}

\preprint{UTF 376}

\btit
{\bf Canonical Quantization of  Photons in a Rindler Wedge }
\etit

\author{Valter Moretti$^\dagger$}

\Idinfn
\begin{center}
(Revised in December 1996)\\
\bigskip
Abstract
\end{center}
\babs \\ 
Photons and thermal photons are studied in the Rindler Wedge 
employing 
Feynman's gauge and   canonical quantization. A Gupta-Bleuler-like
formalism is explicitly implemented. Non thermal 
Wightman functions and related (Euclidean and Lorentzian) 
Green functions are explicitly calculated and their
complex time analytic structure is carefully analyzed
using the Fulling-Ruijsenaars master function.
The invariance of the advanced minus retarded fundamental solution is checked
and a Ward identity discussed.
It is suggested the KMS condition can be implemented to define
thermal states also dealing with unphysical photons.
Following this way, thermal Wightman functions and related 
(Euclidean and Lorentzian) Green functions are built up. Their analytic
structure is carefully examined
employing a thermal master function as in the non thermal case and other
corresponding properties are discussed.
Some subtleties arising dealing with unphysical photons in presence of
the Rindler 
conical singularity are pointed out. 
In particular, a one-parameter family of thermal Wightman and Schwinger
functions with 
the same physical content is proved to exist due to a remaining (non trivial) 
static gauge ambiguity.
A photon version of Bisognano-Wichmann theorem is investigated
in the case of photons propagating in the 
Rindler Wedge employing Wightman functions. Despite of the
found ambiguity in defining Rindler Green functions, 
the coincidence of
$(\beta = 2\pi)$-Rindler Wightman functions 
and Minkowski Wightman functions is proved 
dealing with test functions related to  physical photons
and Lorentz photons.
\eabs 


{V. Moretti: Canonical 
Quantization of Photons...}

\section*{I. INTRODUCTION}


As well-known, the Rindler wedge $W_R$ is defined by the 
inequality $x>|t|$ in a fixed system of rectangular coordinates $(t,x,y,z)$
in  Minkowski space-time. $W_R$
is a globally hyperbolic sub-manifold of Minkowski
space-time. In this paper, we shall consider  $W_R$ as an open set. 
A global coordinate frame  $(\tau,\rho,y,z)$ on
$W_R$ 
is obtained by setting
\beq
x =\rho\cosh\tau \:\:\:\:\mbox{and}\:\:\:\: t = \rho\sinh\tau\:, \label{runo}
\eeq
for $\rho>0$, so
that $x^2-t^2=\rho^2$.\\
Notice that any line $\rho=\rho_0$, $y=y_0$, $z=z_0$
are the trajectory of a uniformly accelerated particle, with
proper acceleration $a=\rho_0^{-1}$ and proper time $s=a\tau$ along the
trajectory. Furthermore, the surfaces $\tau=$ constant are Cauchy
surfaces of $W_R$. The Minkowski metric takes 
the form of the {\em Rindler metric}
\beq
ds^2=-\rho^2d\tau^2+d\rho^2+dx_t^2\:,\nn
\eeq
with $\rho>0$ and
$x_t=(y,z)$.\\
As well-known, 
the Minkowski metric admits 
the timelike Killing field $K=\partial_{\tau}$ in $W_R$.
This vector generates the
isometry $\tau\rightarrow\tau+\tau_0$.
The hypersurface $\rho=0$, i.e., $x^2-t^2=0$
is a {\em Killing event horizon} which {\em bifurcates} \cite{kaywald} in the
transverse two-plane $x=t=0$.\\
We remind that the Rindler metric approximates the metric near the horizon
of a Schwarzschild black hole. In this sense, the physics 
in the Rindler wedge is a toy model
of the physics around a black hole. Thus,
we expect that some result of this paper
can be extended to the Schwarzschild black hole case.\\
In the present paper
we shall study the quantum field theory in
the Rindler region $W_R$ in the case of a photon field
by building up its
Fock representation over the {\em Fulling vacuum}  $|F>$ which is invariant 
under $\tau-$translations 
\cite{full73-7-2850,full77-10-917,unru76-14-870,boul75-11-1404}.
Other authors have studied photon field  or thermal photons
in the Rindler
wedge, directly employing 
the  strength field $F_{\mu\nu}$ instead of the vectorial 
field $A_\mu$ \cite{robavanzo,giapponese}, thus avoiding
gauge related problems, or they have analyzed  particular  problems only
\cite{higuchi}. In this paper we shall develop a more mathematical and 
systematic studying
using the field $A_\mu$.\\  
In {\bf Section I}, we  shall implement a canonical approach to
quantization of the vectorial photon field
using Feynman's gauge,  taking care  to correctly 
deal with the arising unphysical photons. In fact, a Gupta-Bleuler-like
formalism will be  explicitly implemented and the {\em non} Hilbertian
structure of the quantum state space  analyzed.\\
In {\bf Section II}, the Wightman functions will be built up 
within the frame-work of a 
three-smeared distributional formalism. The whole analytic structure
of these functions, the related  Schwinger function,
Feynman propagator 
and the advanced-minus-retarded function  will be analyzed 
employing a  Rindler-time complex {\em master function}
introduced by Fulling and Ruijsenaars for the scalar case 
\cite{full87-152-135}.
In particular, the expected invariance of the advanced-minus-retarded
function will be proved. Finally, a Ward identity will be discussed.\\
In {\bf Section III}, we shall propose a definition of thermal states
in terms of Wightman functions
which uses the KMS condition  also dealing with  unphysical photons.
We shall see that this definition agrees with  all the expected physical
requirements. The thermal Wightman functions will be explicitly built up 
employing the sum over images method and a
thermal master function  analysis  will be implemented. We shall see that  
some gauge ambiguities remain in the definition of these Green functions.
In fact we shall find an one-parameter class of physically equivalent
master (Schwinger, Wightman) functions which, differently from the
scalar case, are defined away from the
conical singularity. This is  due to 
static non trivial 
unphysical terms which affect all the thermal Green functions.
In the case $\be=2\pi$ (absence of  conical singularities) only one
particular Green function defined on the whole Euclidean manifold will arise
from
the above-mentioned class. The Wightman functions, related by analytic
continuation to this special Schwinger function, will give rise to
the local coincidence of the Minkowski vacuum with the $\be=2\pi$
thermal Rindler state.  This coincidence generalizes, in terms of 
Wightman functions, the Bisognano-Wichmann theorem 
\cite{giapponese,sewell,haaglibro}
including 
the photon  field. This  local vacua identity, considered as a Wightman
functions identity, will be
proved  employing physical or Lorentz test wavefunctions. 

\section*{II. PHOTON FIELD QUANTIZATION AND GUPTA-BLEULER FORMALISM
IN A RINDLER WEDGE}
\subsection*{A. INDEFINITE SCALAR PRODUCT}

The first step to quantize a (quasi-)free field theory 
in a  globally hyperbolic space-time consists of the definition of an
appropriate  conserved indefinite scalar product with respect to spatial  
Cauchy 3-surfaces of space-time; this inner product does not 
depend on the particular choice of a Cauchy surface \cite{kaywald,refhiguchi}.
We shall suppose to work in $W_R$ using  coordinates 
$(x^{0},x^{1},x^{2},x^{3}) = (\tau,\rho,y,z)$
defined above. 
It is 
convenient to represent the inner product on the 
$x^{0}=\tau =$ constant spatial surfaces,  they being Cauchy surfaces.
A natural choice of a quantum vacuum (Fulling vacuum in our case)
is obtained by decomposing the field
over normalized  modes which are
imaginary exponentials in the chosen time \cite{birrelldavies,fulling}.
Thus, the creation and destruction operators related
with these  modes define both the quantum vacuum and the corresponding Fock
representation. \\
Let us define the canonical conjugate momentum of the {\em real field}
$A_{\mu}$ in the
case of 
Feynman gauge  as $^{a)}$
\beq
\Pi_{A}^{\mu \nu} :=  \frac{ \partial {\cal L}}{
\partial\partial_{\mu} A_{\nu}}=
\sqrt{-g} [ \lap^{\nu} A^{\mu} - \lap^{\mu} A^{\nu} - g^{\mu \nu}
\lap_{\alpha} A^{\alpha}] = \sqrt{-g}(F^{\mu\nu}-g^{\mu\nu}\lap^\alpha
A_\alpha) \:, \nn
\eeq
where
$\lap_{\mu}$ is the  covariant derivative and
\beq
{\cal L} := -\sqrt{-g} \left[ \frac{1}{4} F_{\mu \nu}F^{\mu \nu} -
\frac{1}{2\alpha} (\lap^{\nu} A_{\nu})^{2}\right]_{\al=1} =
-\frac{\sqrt{-g}}{2}[ \lap_{\mu}A_{\nu} \lap^{\mu}A^{\nu}
+\mbox{(Cov. Deriv.})] \label{lagrangiana}
\eeq
is the Lagrangian of the photon in the {\em Feynman gauge},
$g$ standing everywhere for
the determinant of the complete metric.
Classically, one has to impose the 
{\em Lorentz condition}
\beq
\nabla^{\mu}A_{\mu}=0 \:, \label{gauge}
\eeq
as constraint
on the
solutions of the vectorial Klein-Gordon  equation produced by the
the Lagrangian (\ref{lagrangiana}) as motion equations
\beq
\nabla_{\mu}\nabla^{\mu} A_{\nu}(x)=0 \label{KG}
\eeq
Dealing with generally {\em complex} photon wavefunctions,
the usual canonical conserved indefinite
scalar
product reads \cite{higuchi}: 
\beq
(A,B) = i\: \int_{\Sigma} \: dS
\sqrt{h}\:\:
n_{\mu}\:\:\frac{
\left(A_{\nu}^{\ast} \Pi_{B}^{\mu \nu} - B_{\nu} \Pi_{A}^{\ast \mu \nu}
\right)}{\sqrt{-g}} \label{prodotto2}\:,
\eeq
where $dS := dx^{1}dx^{2}dx^{3}$,
$\Sigma$ is a $x^{0}$ constant Cauchy surface and $n =
-dx^{0}/\sqrt{-(dx^{0},dx^{0})} $ is its normalized,
positive time oriented,
normal vector. Finally, $h$ is the determinant of the  Euclidean 
3-metric $h_{ij}$ induced on $\Sigma$.\\
A simpler 
conserved scalar 
product follows from the {\em Fermi Lagrangian} obtained by dropping the 
classically unimportant total derivative term
in Eq.(\ref{lagrangiana}). The motion equations remain Klein-Gordon Equations
(\ref{KG}).
Dealing with as in the previous case, we obtain a new conserved
inner product:
\beq
(A,B)' = -i\: \int_{\Sigma}  \: dS
\sqrt{h}\:\:
n^{\mu} A_{\nu}^{\ast} \: {\lrup \nabla}_{\mu} \:B^{\nu} 
\:, \nn
\eeq
where $
f \: {\lrup \lap}_{\mu} \:g :=
f \lap_{\mu}g - g \lap_{\mu} f$.
The relation between the two scalar products reads, due to the antisymmetry
of the tensor $D^{\mu\nu}_{AB}$:
\beq
(A,B) - (A,B)' = i\int_{\Sigma} \: dS
\sqrt{h}\:\:
n_{\mu} \nabla_{\nu} D_{AB}^{\mu \nu}\:, \nn  
= 
i\int_{\Sigma} \:\: dx^{1}dx^{2}dx^{3} \sum_{i=1,2,3}
\partial_{i} (\sqrt{-g} D_{AB}^{0 i})
\label{difference}\:,
\eeq
where we defined $D_{AB}^{\mu \nu} :=
  A^{\ast \mu} B^{\nu} - A^{\ast \nu} B^{\mu}$
and used, in our {\em static} coordinates $^{b)}$
\beq
n_{\mu}= \frac{-\delta^{0}_{\mu}}{\sqrt{-g^{00}}}, \:\:\:\:\:
g^{00} = \frac{h}{g}\nn \:\:\:\:\: \mbox{and}\:\:\:\:
dS
\sqrt{h}\:\:
n_{\mu} = -dx^{1}dx^{2}dx^{3} \sqrt{-g} \delta^{0}_{\mu} \nn\:. 
\eeq
The  integral written above becomes a 2-dimensional integral over the edge
$\partial \Sigma$ of $\Sigma$. This vanishes 
provided convenient
boundary conditions on the fields $A_{\mu}$ and $B_{\mu}$ be satisfied
depending
on the behaviour of $\sqrt{-g}$ ($=\rho$ in our case) on this edge and thus
the scalar 
products $(\:,\:)$ and $(\:,\:)'$
coincide. As in the case of a scalar field  \cite{kaywald,fulling},
 it might be possible to 
build up the whole theory 
by considering only real $C^{\infty}$ classical fields
solutions of motion equations with a {\em compact support} on
Cauchy surfaces $\Sigma$. Using  such functions, the two above-mentioned
  scalar
products trivially coincide.
However, this choice (at least dealing with the spinless case)
requires to deal with wavefunctions containing both 
positive and negative frequencies \cite{kaywald,fulling}.
For the time being,
we only assume to deal with positive frequency 
smooth solutions of K-G equations  without to specify further details.
We suppose to deal with the scalar product $(\:,\:)$ only. 
Later, defining Wightman functions, 
we shall employ wavefunctions with spatial compact support 
containing both positive and negative 
frequencies and thus
we shall re-consider the identity $(\:,\:)=(\:,\:)'$.

\subsection*{B. CANONICAL FORMALISM}

Proceeding to the quantization of photon field in $W_R$, using
coordinates $(\tau,\rho,x_{t})$, we have to look
for a decomposition of the {\em real} field $A_{\mu}$  as
\beq
A_{\mu}(x) = \int_{\R^{2}} dk_{t} \:\:\int_{0}^{+\infty} d\omega\:\:
\sum_{\lambda=0}^{3} \{ a_{(\lambda,\omega,k_{t})} 
A_{\mu}^{(\lambda,\omega,k_{t})}(x)  + C.C. \} \label{decomposizione}\:.
\eeq
The positive frequency modes
\beq
A_{\mu}^{(\lambda,\omega,k_{t})}(x) =
A_{\mu}^{(\lambda,\omega,k_{t})}(\rho,x_{t}) e^{-i\omega \tau} \nn
\eeq
must be linearly independent solutions of Klein-Gordon equations (\ref{KG}).
We require the following  
normalization of the modes with respect to the scalar product $(\:,\:)$:
\begin{eqnarray}
(A^{(\lambda,\omega,k_{t})},A^{(\lambda',\omega',k'_{t})})
&=& \eta^{\lambda\lambda'}\: \delta(k_{t}-k'_{t})\:
\delta(\omega-\omega')\label{uno}
\:,\\
(A^{\ast (\lambda,\omega,k_{t})},
A^{\ast (\lambda',\omega',k'_{t})})
&=& -\eta^{\lambda\lambda'}\: \delta(k_{t}-k'_{t})
\:\delta(\omega-\omega')
\label{due}
\:,\\
(A^{\ast (\lambda,\omega,k_{t})},
A^{(\lambda',\omega',k'_{t})})
&=& 0
\label{tre}
\:,
\end{eqnarray}
where $\eta_{\mu \nu} \equiv \eta^{\mu\nu}\equiv \mbox{diag}\:(-1,1,1,1,)$
$^{c)}$. From 
Eq.(\ref{uno}), it arises: 
\beq
(A,A') 
= \int_{\R^{2}} dk_{t} \:\:\int_{0}^{+\infty} d\omega\:\:
\sum_{\lambda=0}^{3} 
a^{\ast}_{(\lambda,\omega,k_{t})} 
a'_{(\lambda,\omega,k_{t})} \eta^{\lambda \lambda'}\:,\label{scalarproduct}
\eeq
where
$A_{\mu}$ and $A'_{\mu}$ are
(generally complex) {\em positive frequency} photon wavefunctions.
The Fourier coefficients $a_{(\lambda,\omega,k_{t})}$ 
are such that the corresponding positive frequency wavefunction $A$ 
results to be 
 smooth $^{d)}$.
These coefficients, understood as functions of the variables
$\omega$ and $k_t$, define one-particle quantum states $|\Psi_A>$.\\
Holding the  
normalization relations (\ref{uno}), (\ref{due}) and (\ref{tre}),
it simply follows from Eq.(\ref{decomposizione}):
\begin{eqnarray}
a_{(\lambda,\omega,k_{t})}&=& \eta_{\lambda \lambda'}
(A^{(\lambda',\omega',k'_{t})}, A) \nn\:, \\
a^{\ast}_{(\lambda,\omega,k_{t})}&=& - \eta_{\lambda \lambda'}
(A^{\ast(\lambda',\omega',k'_{t})}, A) \nn\:.
\end{eqnarray}
We have to  interpret these coefficients as operators to
quantize.
As usually, the equal time {\em canonical commutations rules}
of the operator $\hat A$ and its  conjugate momentum  $\hat \Pi$
imply the bosonic {\em commutations rules} of the  
operators $\hat a$ and  $\hat a^\dagger$. They read
respectively:
\beq
[ \hat A_{\mu}(x), n_{\alpha} \hat\Pi^{\alpha \nu}(x')]_{\tau=\tau'} =
i \delta^{\nu}_{\mu}  \:\: 
\delta(\rho-\rho') \:\: \delta(x_{t}-x'_{t}) \: \mbox{I} \label{CCR}
\eeq
{\em [The remaining  (independent)
equal time commutators vanish]}\\
where  $n:= - d\tau /\sqrt{-(d\tau,d\tau)}$,
and
\beq
[\hat a_{(\lambda,\omega,k_{t})},\hat a^{\dagger}_{(\lambda',
\omega',k'_{t})}]&=&
\eta_{\lambda\lambda'}\: \delta(k_{t}-k'_{t})
\:\delta(\omega-\omega') \: \mbox{I} \label{commut}
\eeq 
{\em [The remaining (independent)
commutators vanish]} \\
We expect that not all the
one-particle quantum states are representable 
by 
smooth positive frequency wavefunctions. This should hold
only for states belonging in  a
linear manifold ${\cal M}$ supposed to be {\em dense} (in some topology) 
in the whole one-particle quantum states space.
 We expect that  one-particle quantum states
space ${\cal H}$ can be represented as an
{\em algebraic} tensorial product $^{e)}$
${\cal H} := \C^4 \otimes {\cal H}_0 $
where $\R_+:=[0,+\infty)$, ${\cal H}_0 $ being 
$L^2(\R^2 {\bf \times} \R_+) $ or a proper closed subspace of this.
We can write:
\beq
|\Psi> \equiv (\Psi_0(k_t,\omega),\Psi_1(k_t,\omega),\Psi_2(k_t,\omega),
\Psi_3(k_t,\omega)) \:\:\:\: k_t \in \R^2,\:\:\: \omega \in \R_+
\label{rappsi}
\eeq
Finally, ${\cal H}$ has to be endowed with a scalar product
compatible with 
the above-written  commutation relations. This reads:
\beq
<\Psi|\Psi'> := \sum_{\lambda, \lambda'=0}^{3}\eta^{\lambda\lambda'}
\int_{\R} dk_t \int_{\R_+}d\omega
\:\:  
\Psi^{\ast}_{\lambda}(k_t,\omega)\Psi'_{\lambda'}(k_t,\omega)
\nn\:.
\eeq
The space ${\cal H}$ {\em cannot} 
be considered a  properly defined Hilbert space
due to presence of the {\em indefinite} matrix $\eta$.
This matrix  appears due to the  
 unphysical degrees of freedom represented by non transverse photons
 necessary in order to deal with a gauge constraint 
 explicitly covariant as the Lorentz condition Eq.(\ref{gauge}).
The problem is the same as in Minkowski coordinates quantization.\\
The positive frequency 
wavefunction related with a {\em generic} vector $|\Psi> \in {\cal M}$ 
is the positive frequency smooth  function:
\beq
A_\mu(x) = \int_{\R} dk_t \int_{\R_+}d\omega
\:\:\sum_{\lambda=0}^3\:
\Psi_{\lambda}(k_t,\omega) A^{(\lambda,\omega,k_t)}(x) 
\label{wavefunctions}\:,
\eeq
and thus  the coincidence of the two scalar product
$(A,B)$ and $<\Psi_A|\Psi_B>$ results to be evident.\\
The inverse formula of Eq.(\ref{wavefunctions}),  
following from the normalization relations of the modes Eq.s (\ref{uno}),
(\ref{due}) e (\ref{tre}), holds in the same space ${\cal M}$: 
\beq
\Psi_{\lambda}(k_t,\omega) = -( A^{(\lambda,\omega,k_t)}, A) = \nn
\eeq
\beq
=-i\int_{\Sigma} \: dS
\sqrt{h}\:
n_{\mu}\:
\left[ A_{\nu}^{\ast(\lambda,\omega,k_t)} 
(F_{A}^{\mu \nu}- g^{\mu\nu}\nabla_\alpha A^\alpha
) - A_{\nu} ( F_{ A^{\ast (\lambda,\omega,k_t)} }^{\mu \nu}
- g^{\mu\nu}\nabla_\alpha A^{\ast (\lambda,\omega,k_t)\alpha })\right] 
\nn\eeq
or, provided $A$ vanish on the edge of $\Sigma$, it reads:
\beq
\Psi_{\lambda}(k_t,\omega) = -( A^{(\lambda,\omega,k_t)}, A)' =
i \int_{\Sigma}  \: dS
\sqrt{h}\:\:
n^{\mu} A_{\nu}^{\ast(\lambda, k_t,\omega)} \: {\lrup \nabla}_{\mu} \:A^{\nu}
\nn\:.
\eeq
The  relations (\ref{commut}) 
have to be more correctly  understood  as:
\beq
[\hat a_{\Psi},\hat a^{\dagger}_{\Psi'}] 
\:\:\: = \:\:\: <\Psi|\Psi'>\:\mbox{I}\:\:\:\: ( = (A_\Psi,A_{\Psi'})
\:\mbox{I}\:\:\:\:
\mbox{if}\:\:\:\: |\Psi>,|\Psi'>\in {\cal M})\nn
\eeq
{\em [The remaining (independent)
commutators vanish]}\\
where $\hat a_\Psi$ and $\hat a^\dagger_\Psi$, when $|\Psi>\in {\cal M}$, are 
interpreted as 
\begin{eqnarray}
\hat a_{\Psi_A}&=&
(A, \hat A) \nn\:\:\:\mbox{(antilinear in A)} \:,
\label{aleph} \\
\hat a^{\dagger}_{\Psi_A}&=& -
(A, \hat A) \nn\:\:\:\:\:\mbox{(linear in A)}\:.
\label{beth}
\end{eqnarray}
$\hat A$ has to be interpreted as an {\em operator
valued distribution} working  on
smooth positive frequency
photon wavefunctions $A$. \\
These identities make sense dealing with an appropriate invariant 
linear manifold dense 
(in some topology) in the
 symmetrized Fock-like space ${\cal F}({\cal H})_s$, {\em algebraically}
built up as a normal symmetrized Fock space,
 $\hat a$ and $\hat a^{\dagger}$ being  destruction
and creation
operators. The {\em Fulling vacuum} $|F>$ is defined as:
\beq
\hat a_{\Psi} |F>=0 
\:\:\:\: |\Psi> \in {\cal H} \nn
\:.
\eeq
By the normalization relations and, in particular, because of the trivial time
dependence of the modes, the Rindler Hamiltonian of the
photons results to be
 diagonal if written in terms of 
$\hat a_{(\lambda,\omega,k_{t})}$
and $\hat a^{\dagger}_{(\lambda',\omega',k'_{t})}$, the spectral 
parameter
being $\omega$. Using the 
normal order prescription, we have:
\beq
:{\hat H}: = :\int_\Sigma dS (\sqrt{-g}\:
 n_\sigma \hat\Pi^{\sigma\nu} n_\lambda
\partial^{\lambda}\hat A_\nu - \hat{\cal L}) : =
 \int dk_t \: d\omega \: \omega \:
\sum_{\lambda=0}^3 \eta^{\lambda\lambda}\: 
\hat a^{\dagger}_{(\lambda,\omega,k_{t})} \hat a_{(\lambda,\omega,k_{t})}
\nn\:.
\eeq
Thus we can consider the
 quanta generated by $\hat a^{\dagger}_{(\lambda',\omega',k'_{t})}$ as
 defined Rindler-energy particles. 
However, there arise particles of
{\em  negative} norm and energy due the 
anomalous commutation rule of $\hat a_{\lambda=0}$ and
$\hat a^{\dagger}_{\lambda=0}$
as in the 
Minkowskian case $^{f)}$.
We expect that a 
{\em Gupta-Bleuler}-like formalism (see
\cite{itzykson} for example) can be used in order to deal more correctly
with the Feynman gauge.\\

\subsection*{C. NORMAL MODES AND ONE-PARTICLE SPACE}

Let us seek a set of normal modes solutions of Klein-Gordon equation and
satisfying the constraints in Eq.s (\ref{uno}), (\ref{due}), (\ref{tre}).  
We report the results  and some comments here.
All the calculations are contained in 
{\bf Appendix A}. \\
We start with the  independent 
modes suggested by Higuchi, Matsas and Sudarsky
in \cite{higuchi}.
\begin{eqnarray}
A_{\mu}^{(I,\omega,k_{t})} &\equiv & C^{(I,\omega,k_{t})}(0,0,k_{z}
\phi,-k_{y}\phi) \label{auno}\:,\\
A_{\mu}^{(II,\omega,k_{t})} &\equiv & C^{(II,\omega,k_{t})}
(\rho\partial_{\rho}\phi,-i\frac{\omega}{\rho}\phi,0,0) \label{adue}\:,\\
A_{\mu}^{(G,\omega,k_{t})} &\equiv & C^{(G,\omega,k_{t})}(-i\omega \phi,
\partial_{\rho}\phi ,ik_{y}
\phi,ik_{z}\phi) \equiv C^{(G,\omega,k_{t})}\label{atre} \: \partial_\mu \phi
\:,\\
A_{\mu}^{(L,\omega,k_{t})} &\equiv & C^{(L,\omega,k_{t})}(0,0,k_{y}
\phi,k_{z}\phi) \label{aquattro}\:,
\end{eqnarray}
where the coefficients $C$ are normalization constants, and the field 
$\phi = \phi^{(\omega,k_{t})}(x)$
is the
 mode solution of {\em scalar}   Klein-Gordon equation in $W_R$:
\beq
\phi^{(\omega,k_{t})}(x) = K_{i\omega}(k_{\perp}\rho) 
e^{i(k_{t}x_{t}-\omega\tau)} \:. \label{phi} 
\eeq
$K_{\nu}(z)$ is a well-known MacDonald function of imaginary
index
\cite{tricomi}, $k_{\perp}:=\sqrt{k_{y}^{2}+k_{z}^{2}}$ and
$k_{t}x_{t}:= k_{y}y+ k_{z}z$.
The Klein-Gordon equation reads:
\beq
(-\frac{1}{\rho} \partial_{\tau}^{2} + \partial_{\rho} \rho \partial_{\rho}
+ \rho \nabla_t^2) \phi = 0 \nn\:,
\eeq
and the solution $\phi= \phi^{(\omega,k_{t})}$ also satisfies:
\beq
\partial_{\tau}^{2} \phi = -\omega^{2}\phi \:\:\:\:\mbox{and}\:\:\:\:
\nabla_t^2 \phi :=
\sum_{\alpha= y,z}\partial_{\alpha}^{2} \phi = -k_{\perp}^{2} \phi \nn
\:.
\eeq
It can be simply 
proved that the modes $I$ and $II$ satisfy the Lorentz  
condition (\ref{gauge}).  \\
The mode $G$ is proportional to  
$\partial_{\mu}\phi$
and thus it is a {\em pure gauge mode}; note that this also satisfies the
Lorentz condition because $\phi$ is a solution of
scalar Klein-Gordon equation. The mode $L$ does not satisfy the
gauge condition. 
Using the inner product $(\:,\:)$, 
one finds 
the mode $I$ to be  normal to the mode $II$,
furthermore the linear space spanned 
by the unphysical modes $G$ and $L$ results to be  normal to
the modes $I$ and $II$.
Departing from the work 
\cite{higuchi}, we follow
 a different choice of unphysical modes in order to have
a complete set of  normal to each other modes.
We define new modes, considering two convenient  linear combinations of
 unphysical modes $G$ and $L$:
\begin{eqnarray}
A_{\mu}^{(1,\omega,k_{t})} &\equiv & C^{(1,\omega,k_{t})}(0,0,k_{z}
\phi,-k_{y}\phi) \label{buno}\:,\\
A_{\mu}^{(2,\omega,k_{t})} &\equiv & C^{(2,\omega,k_{t})}
(\rho\partial_{\rho}\phi,-i\frac{\omega}{\rho}\phi,0,0) \label{bdue}\:,\\
A_{\mu}^{(3,\omega,k_{t})} &\equiv & C^{(3,\omega,k_{t})}(-i\omega \phi,
\partial_{\rho}\phi,
0,0) \label{btre}\:,\\
A_{\mu}^{(4,\omega,k_{t})} &\equiv & C^{(4,\omega,k_{t})}(0,0,ik_{y}
\phi,ik_{z}\phi) \label{bquattro}\:.
\end{eqnarray} 
Following the calculations of {\bf Appendix A} we may define  normalized modes 
$A_{\mu}^{(\lambda,\omega,k_{t})}$:
\begin{eqnarray}
A_{\mu}^{(0,\omega,k_{t})} &\equiv & \frac{\sqrt{\sinh
\pi \omega}}{2 \pi^{2} k_{\perp}} 
\: (-i\omega \phi,
\partial_{\rho}\phi,
0,0) \equiv A_\mu^{(G,\omega,k_t)}-iA_\mu^{(L,\omega,k_t)} \label{ezero}\:,\\
A_{\mu}^{(1,\omega,k_{t})} &\equiv & \frac{\sqrt{\sinh \pi 
\omega}}{2 \pi^{2} k_{\perp}} \:   (0,0,k_{z}
\phi,-k_{y}\phi) \label{euno}\:,\\
A_{\mu}^{(2,\omega,k_{t})} &\equiv & \frac{\sqrt{\sinh \pi \omega}}{2 
\pi^{2} k_{\perp}}\:
(\rho\partial_{\rho}\phi,-i\frac{\omega}{\rho}\phi,0,0) \label{edue}\:,\\
A_{\mu}^{(3,\omega,k_{t})} &\equiv & \frac{\sqrt{\sinh \pi \omega}}{2 
\pi^{2} k_{\perp}}\:(0,0,ik_{y}
\phi,ik_{z}\phi) \equiv iA_\mu^{(L,\omega,k)} \label{etre}\:.
\end{eqnarray}
Using these modes, the normalization relations Eq.s (\ref{uno}), 
(\ref{due}),
(\ref{tre}) are satisfied.

To conclude, we are able to suggest
 a possible definition of the space ${\cal M}$ and the one-particle space
${\cal H}$. However, we shall not study  this topic in deep.
 Let us consider the set $S$ of the $C^\infty$ {\em real} wavefunctions 
solutions of the vectorial  K-G equation
with spatial {\em compact support} at $|\tau|<+\infty$ $^{g)}$
and such that their transverse Fourier transform vanishes with order
$|k_t|^n$
$n \geq 1 $ as $k_t\rightarrow 0$. They are, for example,
transverse coordinate  Laplacians of $C^\infty$
 compact support K-G solutions.
 The required condition,
passing to the Fourier decoposition through Eq.(\ref{psia}) (see below)
 cancels against the divergent factor $k_\perp^{-1}$in the modes  and assures
a finite $L^2$ norm (see below) $^{h)}$.
The following decomposition arises (note the presence of negative
 frequencies):  
\beq
A_\mu(x) = \int_{\R} dk_t \int_{\R_+}d\omega
\:\:\sum_{\lambda=0}^3\:
\Psi_{\lambda}(k_t,\omega) A_\mu^{(\lambda,\omega,k_t)}(x) +
\sum_{\lambda=0}^3\:
\Psi^{\ast}_{\lambda}(k_t,\omega) A_\mu^{\ast(\lambda,\omega,k_t)}(x)
\label{doublewavefunctions}\:,
\eeq
where 
\beq
\Psi_{\lambda}(k_t,\omega) = -( A^{(\lambda,\omega,k_t)}, A)=
-( A^{(\lambda,\omega,k_t)}, A)' \label{psia}\:,
\eeq
because the boundary terms vanish due to
compact support of $A$. Employing the previously found normal modes 
and standard properties of MacDonald's functions \cite{tricomi} it is 
possible to prove that these scalar products are finite. Similarly 
it is possible to obtain ($\lambda = 0,1,2,3$): 
\beq
\int_{\R^2{\bf \times} \R_+} | \Psi_\lambda(k_t,\omega)|^2 
dk_t d\omega < +\infty \:, \nn
\eeq
when $A\in S $. 
We define ${\cal S} := <S>_{\R}$, i.e., the {\em real} linear space spanned
by the set $S$. We can define ${\cal M}$ as the {\em complex} linear
space ${\cal M} := <M>_{\C}$, where $M$
is the set of states defined by the left hand side of Eq.(\ref{psia}) when
$A\in S$. The  positive frequency 
wavefunctions of the state $|\Psi>\in {\cal M}$ is
written in Eq.(\ref{wavefunctions}). 
The one photon Hilbert space is thus defined as ${\cal H} = 
\overline{{\cal M}} =\C^4 \otimes_T{\cal H}_0 $. The closure as well as
 the topological
tensorial product will be defined employing a certain topology specified 
in the following section.

\subsection*{D. GUPTA-BLEULER FORMALISM IN A RINDLER SPACE}

Following the Minkowskian theory,
the  quantum states space of the whole theory
must be   formally defined
as the space of vectors
$|\Psi> \in {\cal F}({\cal H})_s$ satisfying the Lorentz constraint:
\beq
\nabla^{\mu} \hat A^{+}_{\mu}(x)|\Psi> = 0 \:, \label{quantumgauge}
\eeq
where $\hat A^{+}_{\mu}$ is the part of the  operator $\hat A$
containing only destructor operators. 
Eq.(\ref{quantumgauge}) is equivalent to:
\beq
(\hat a_{(3,k_{t},\omega)}-\hat a_{(0,k_{t},\omega)})
|\Psi> =0 \:\:\:\:\:\: \mbox{for all the} \:\: k_{t},\omega \label{GB}\:.
\eeq
This equation defines the physical quantum states exactly as in the case
of the Minkowski theory. 
Furthermore, it can be simply  proved that:
\beq
\lap^{\mu} A_{\mu}^{(\alpha,\omega,k_{t})} = 0 \:\:\:\: \mbox{for}\:\: \alpha=
1,2 \:. \label{za}
\eeq
These modes defines {\em real} particles, endowed with
 a positive norm and  a positive energy.
They are the transverse photons,
namely the two physical degrees of freedom of Rindler photons.\\
The {\em Gupta-Bleuler} formalism can be employed
as in  flat coordinates.
Eq.(\ref{GB}) reads in a non formal representation  
(where the indices of destructor operators are referred to
the notation in Eq.(\ref{rappsi})):
\beq
(\hat a_{(\Phi,0,0,0)}-\hat a_{(0,0,0,\Phi)})
|\Psi> =0 \:\:\:\:\:\: \mbox{for all} \:\: \Phi\in {\cal H}_0\:.
\label{abcd}
\eeq
The space of these vectors will be termed ${\cal F}({\cal H}_L)_s$.
The {\em one particle} space ${\cal H}_L$ can be defined introducing the 
operators:
\beq
\hat\alpha_\Phi := \frac{1}{\sqrt{2}}
(\hat a_{(\Phi,0,0,0)}-\hat a_{(0,0,0,\Phi)}) \:\:\:\: \mbox{and}\:\:\:\: 
\hat\beta_\Phi := \frac{1}{\sqrt{2}}
(\hat a_{(\Phi,0,0,0)}+\hat a_{(0,0,0,\Phi)}) \nn
\eeq
Thus
Eq.(\ref{abcd}) reads simply:
\beq
\hat\alpha_\Phi |\Psi> =0 \:\:\:\:\:\: \mbox{for all the} \:\: \Phi\in
{\cal H}_0\:,\nn
\eeq
and  ${\cal H}_L$ contains one-particle states defined by the
creation operators $\hat{a}^\dagger_1$, $\hat{a}^\dagger_2$, 
$\hat \alpha^\dagger =\hat{\beta}^\star$ only (the adjoint 
$\hat{\beta}^\star$ is defined below).\\
One can trivially obtain the form of the wavefunction
of $|\Psi>\in {\cal H}_L$ passing to the base of the modes
$A^{(1)}$, $A^{(2)}$, $A^{(G)}$, $A^{(L)}$ previously introduced.
This reads:
\beq
A^L_\mu(x) = \int_{\R^2} dk_t \int_{\R_+}
d\omega \: \sum_{\lambda= 1,2,G} \Psi_\lambda(k_t,\omega) A^{(\lambda,\omega,
k_t)}_\mu(x)\:.\nn
\eeq 
These wavefunctions satisfy the Lorentz condition in Eq.(\ref{gauge}).\\ 
We define the {\em physical Fock space} 
${\cal F}({\cal H}_P)_s$
by requiring the total absence of  
{\em unphysical} photons:
\beq
\hat a_{(\Phi,0,0,0)}|\Psi> = \hat a_{(0,0,0,\Phi)}
|\Psi> = 0 \:\:\:\:\:\: \mbox{for all}
\:\: \Phi\in {\cal H}_0\:.\nn
\eeq
The wavefunctions of the states of this space read:
\beq
A^P_\mu(x) = \int_{\R^2} dk_t \int_{\R_+}
d\omega \: \sum_{\lambda= 1,2} \Psi_\lambda(k_t,\omega) A^{(\lambda,\omega,
k_t)}_\mu(x)\nn
\eeq
Obviously, it holds ${\cal F}({\cal H})_s\supset {\cal F}({\cal H}_L)_s 
\supset
{\cal F}({\cal H}_P)_s$.\\
It is possible to define a metrical topology compatible with the physics
of the system.
Following the Minkowski Gupta-Bleuler formalism,
we define a new, {positive defined}, scalar product of $|\Psi>$
and $|\Psi'>$  $\in {\cal H}$ by
\beq
<\Psi/\Psi'> := \sum_{\lambda,\lambda'=0}^{3}\delta^{\lambda\lambda'}
\int_{\R} dk_t \int_{\R_+}d\omega
\:\:
\Psi^{\ast}_{\lambda}(k_t,\omega)\Psi'_{\lambda'}(k_t,\omega)
\nn\:.
\eeq
We shall call this unphysical scalar product the {\em Euclidean} scalar
product.
Employing  this we may make ${\cal H}= \C^4 \otimes_T
{\cal H}_0$ a correctly defined 
{\em Hilbert space}.
${\cal H}_0$ results to be $L^2(\R^2 {\bf \times} \R_+)$ or a proper closed subspace
of this obtained imposing $\overline{{\cal M}} = \C^4\otimes_T {\cal H}_0$.
We can correctly define the Fock space ${\cal F}({\cal H})_s$ as a
Hilbert space, too. Finally, we can define the spaces 
${\cal F}({\cal H}_L)_s$ and ${\cal F}({\cal H}_P)_s$
as  closed  subspaces by taking the topological closure of the 
corresponding algebraically defined linear manifolds.  
Using the norm related with the Euclidean scalar product, i.e., the Euclidean
norm, 
 we can also define topological tools on  the space
of operators, in particular we can define the adjoint 
(including its domain) of an operator
$\hat{\cal O}$ represented by the symbol
  $\hat{\cal O}^\star$.  Then we introduce on 
${\cal F}({\cal H})_s$ the  limited operator $M$ as
the only operator satisfying $^{i)}$:
\beq
M |F> = |F>, \:\:\:\:\: MM = I, \:\:\:\:\: M^\star=M \nn
\eeq
and furthermore:
\beq
M \hat{a}_\lambda = \eta^{\lambda\lambda} \hat{a}_\lambda M 
\:\:\:\:\:\: (\:\mbox{and thus} \:\:\:\:\: 
M \hat{a}^{\dagger}_\lambda = \eta^{\lambda\lambda} 
\hat{a}^{\dagger}_\lambda M) \nn
\:.
\eeq
Now, we can define the the {\em physical} scalar
product, by the continuous sesquilinear
form:
\beq
<\Psi|\Psi'> := < \Psi/ M /\Psi'> \label{scalar2}\:.
\eeq
Employing this definition, one has to define
the adjoint with respect to the physical scalar product (including the 
definition of the domain of the adjoint) 
as
\beq
\hat{{\cal O}}^\dagger = M \hat{{\cal O}}^{\star}M\:.\nn
\eeq
In particular, it holds:
\beq
\hat{a}_\lambda^\star= \eta_{\lambda\lambda}\hat{a}_\lambda^{\dagger},
\:\:\:\hat{\alpha}^\dagger=\hat{\beta}^\star,\:\:\:
\hat{\beta}^\dagger=\hat{\alpha}^\star
\:\:\:\:
\mbox{and}\:\:\:\: 
M^\dagger= M^\star= M \:.\nn
\eeq
Starting from  formulae obtained above, 
it can be simply proved that the following statements hold:
\beq
\mbox{if} \:\:\:\:\: |\Psi>, \:|\Psi'> \in {\cal F}({\cal H}_P)_s\:\:\:\:\:
\mbox{then}\:\:\:\:\: <\Psi|\Psi'> = <\Psi/\Psi'> \nn \:,
\eeq
\beq
\mbox{if} \:\:\:\:\: |\Psi>, \:|\Psi'> \in {\cal F}({\cal H}_L)_s \:\:\:\:\:
\mbox{then}\:\:\:\:\: <\Psi|\Psi'> = <\Psi/P_P/\:\Psi'> \nn \:,
\eeq
where $P_P$ is the  normal projector onto the (closed) physical
Fock space 
${\cal F}({\cal H}_P)_s$. Working in the space ${\cal F}({\cal H}_L)_s$,
only the  physical part of the state contributes to the
physical scalar products.
Thus, we have found exactly the same features which appear
in  Minkowski theory \cite{itzykson}.
In particular, a necessary condition  to consider an operator 
$\hat{{\cal O}}$ 
physically sensible, i.e., 
a {\em gauge invariant observable}, 
consists of the requirement
\beq
<\Psi|\hat{{\cal O}}|\Psi'> = <\Psi/ P_P\: \hat{{\cal O}}
\: P_P/ \Psi'> \nn
\:,
\eeq
for all the vectors $|\Psi>, |\Psi'>\in {\cal F}({\cal H}_L)_s$ 
which make sensible the
left hand side of this identity. 

\section*{III. WIGHTMAN FUNCTIONS AND RELATED GREEN FUNCTIONS}

\subsection*{A. WIGHTMAN FUNCTIONS, RELATED GREEN FUNCTIONS AND THEIR
ANALYTIC STRUCTURE}

In this section we shall calculate following a quite rigorous 
way the {\em Wightman functions} of Fulling
vacuum
for the field $\hat A_{\mu}$.
They are the distributional kernel involved calculating
\beq
<F| (A,\hat{A}) (A',\hat A)|F> =\nn
\eeq
\beq
=\int_\Sigma dS \sqrt{h(x)}\: n^\nu \int_\Sigma dS'\sqrt{h(x')}
\: n'^{\nu'}\: A^\mu (x) A'^{\mu'}(x') \:{\lrup\nabla}_{\nu}
{\lrup\nabla}_{\nu'}\:
<F| \hat A_{\mu}(x) \hat A_{\mu'}(x')|F> \label{wightmandef}
\eeq
where $A$ and $A'$ belong to ${\cal S}$. Notice that, employing such
wavefunctions, $(\:,\:) = (\:,\:)'$. We shall use only the notation $(\:,\:)$
for sake of simplicity.
We stress that the left hand side of Eq.(\ref{wightmandef}) is defined 
 independently on the right hand side:
$<F| \hat A_{\mu}(x) \hat A_{\mu'}(x')|F>$ is  defined just
to make sensible Eq.(\ref{wightmandef}) in a 
distributional-like sense.\\
Alternatively one could try to define the Wightman functions by
using a {\em four smeared} formalism (see {\bf Appendix C}).
In this paper we prefer to use the {\em three smeared formalism}
based on
 Eq.(\ref{wightmandef}).\\
Eq.s (\ref{aleph}) and (\ref{beth}) lead us to
\beq
(A,\hat A) = \hat{a}_{\Psi_A} - \hat{a}^{\dagger}_{\Psi_A}\nn\:,
\eeq
holding when $A, A' \in {\cal S}$ are {\em real}  and thus
$|\Psi_A>, |\Psi_{A'}>\in {\cal M}$ are obtained employing 
Eq.(\ref{psia}).
Substituting this in the left hand side of Eq.(\ref{wightmandef}), 
it arises $^{j)}$
\beq
<F| (A,\hat A) (A',\hat A)|F> = -<F|\hat{a}_{\Psi_A}\hat{a}^{\dagger}_{\Psi_A}
|F> = -<\Psi_A|\Psi_{A'}>=\nn
\eeq
\beq
= -\int_{\R^2}dk_t \int_{\R^+} d\omega  \sum_{\lambda\lambda'}
\eta^{\lambda\lambda'}\:
\Psi^\ast_{\lambda A}(k_t,\omega)\:\Psi_{\lambda'A'}(k_t,\omega) =\nn
\eeq 
\beq
= -\sum_{\lambda \lambda'} \eta_{\lambda\lambda'}
\int_{\R^2}dk_t \int_{\R^+} d\omega \:
\int_{\Sigma}  \: dS
\sqrt{h}\:\:
n^{\mu} A_{\nu}^{(\lambda,k_t,\omega)} \: {\lrup \nabla}_{\mu} \:
A^{\nu}
\int_{\Sigma}  \: dS'
\sqrt{h}\:\:
n^{\mu'} A_{\nu'}^{\ast(\lambda',k_t,\omega)} \: {\lrup \nabla}_{\mu'}
\:A'^{\nu'}\:.\nn
\eeq
It is possible  to change the order of 
the integrals in the right hand side of the 
 equation written above by introducing an
$\ep-${\em prescription} in the time variable  appearing into normal modes.
Thus we obtain in a distributional sense (namely taking the limit 
$\ep \rightarrow 0^+$ at the end of calculation):
\beq
<F| (A,\hat A) (A,\hat A')|F> =\nn
\eeq
\beq
=-\int_\Sigma dS \sqrt{h(x)}\: n^\nu \int_\Sigma dS'\sqrt{h(x')}
\: n'^{\nu'}\: A^\mu (x) A^{\mu'}(x') \:\lrup{\nabla}_{\nu}
\lrup{\nabla}_{\nu'}\:
<F| \hat A_{\mu}(x) \hat A_{\mu'}(x')|F> \nn \:,
\eeq
where:
\beq
<F| \hat A_{\mu}(x) \hat A_{\mu'}(x')|F> 
(= <F| \hat A_{\mu}(x) \hat A_{\mu'}(x')|F>_\ep) =\nn
\eeq
\beq
:=\int_{\R^{2}} dk_{t}\int_{0}^{+\infty} d\omega \:\sum_{\lambda\lambda'}
\eta_{\lambda\lambda'}
\:\: A_{\mu}^{(\lambda,\omega,k_{t})}(\rho,x_t)
\:\: A_{\mu'}^{\ast (\lambda,\omega,k_{t})}(\rho',x'_t) \: e^{-i\omega
(\tau-\tau'-i\ep)}\:. \nn
\eeq
Using less rigor, we can obtain the same result by expanding
the field operators which appear in the  formal object
$<F| \hat A_{\mu}(x) \hat A_{\mu'}(x')|F>$ over the normal modes and
introducing the $\ep-$prescription  by hand.\\
Using the modes in Eq.s (\ref{ezero}), 
(\ref{euno}),  (\ref{edue}) and 
(\ref{etre}), the expression (\ref{phi}) of the field $\phi$ and substituting
all  them in the equation above, one finds the following 
integral decomposition:
\beq
W^{+}_{\mu\mu'}(x,x') \:\:\:(= W^{+}_{\mu\mu'}(x,x')_\ep)
:=
<F| \hat A_{\mu}(x) \hat A_{\mu'}(x')|F>=\nn
\eeq 
\beq
=\frac{1}{4\pi^{4}} \:
\int_{\R^{2}} dk_{t} \frac{e^{ik_{t}(x_{t}-x'_{t})}}{k_\perp^2} 
\:D_{\mu\mu'}\:\int_{0}^{+\infty} d\omega\:\sinh \pi\omega
\:
K_{i\omega}(k_{\perp}\rho) K_{i\omega}(k_{\perp}\rho')\:
e^{-i\omega(\tau-\tau'-i\ep)}\:,\label{F}
\eeq
the operator $D_{\mu\mu'}$ is defined as
\begin{eqnarray}
D_{\tau\tau'}&:=& -\partial_{\tau}\partial_{\tau'}+\rho\partial_{\rho}
\rho'\partial_{\rho'} \:, \nn\\
D_{\rho\rho'}&:=& \frac{1}{\rho\rho'} 
[\partial_{\tau}\partial_{\tau'}-\rho\partial_{\rho}
\rho'\partial_{\rho'}] \nn\:,\\
D_{\tau \rho'}&:=&\frac{1}{\rho'}\: [\partial_{\tau'}\rho \partial_{\rho}
-\partial_{\tau}\rho' \partial_{\rho'}] \:, \nn\\
D_{\rho \tau'}&:=&\frac{1}{\rho}\: [\partial_{\tau}\rho' \partial_{\rho'}
-\partial_{\tau'}\rho \partial_{\rho}] \:, \nn\\
D_{yy'}&:=& D_{zz'}= k_\perp^2 (= k_t^2) \nn
\end{eqnarray}
all the remaining terms vanish.\\
We shall explicitly 
calculate the integrals written above employing an indirect method (details
are reported in {\bf Appendix B}).\\
We needs some preliminary definitions and results. \\
Let us define the quantity $\alpha$ by:
\beq
\cosh\alpha(\rho,\rho',x_t-x'_t) 
=\frac{\rho^2+\rho^{'2}+|x_t-x^{'}_t|^2}{2\rho\rho^{'}}
= \frac{1}{2}(\frac{\rho}{\rho'}+\frac{\rho'}{\rho}+
\frac{|x_t-x^{'}_t|^2}{\rho\rho'})
\label{alpha}
\eeq
and let us remember the form of the Wightman function on the Fulling vacuum 
of a {\em massless}
scalar field propagating in $W_R$ (see \cite{mova} and Ref.s therein):
\beq
W^{+}(x,x') = W^{+}(\tau-\tau',\rho,\rho',x_{t}-x'_{t})=\nn
\eeq
\beq 
= \frac{1}{4\pi^{4}} \:
\int_{\R^{2}} dk_{t}e^{ik_{t}(x_{t}-x'_{t})}\:
\int_{0}^{+\infty} d\omega\:\sinh \pi\omega
\: K_{i\omega}(k_{\perp}\rho) K_{i\omega}(k_{\perp}\rho')\:
 e^{-i\omega(\tau-\tau'-i\ep)} =\nn
\eeq
\beq
= \frac{1}{4\pi^2}\frac{\al}{\rho
\rho^{'}\sinh\al}\frac{1}{\al^2 -(\tau-\tau^{'}-i\ep)^2}\:.\label{W}
\eeq
 The integrand in  the latter formula differs from the integrand in 
Eq.(\ref{F}) due to
 the absence of  the factor $k_\perp^{-2}$ and the operator
$D_{\mu\mu'}$, only. 
Remind that it formally holds in $\R^{2}$:
\beq
2\pi \ln \frac{|x_t|}{\mu_0} = \int_{\R^{2}} dk_t 
\frac{e^{ ik_t x_t}}{k_t^{2}} \:\:\:\: (\mbox{where}\:\:\:\: \mu_0 := 
\int_0^1 du \frac{1-J_0(u)}{u}-\int_1^{+\infty} du 
\frac{J_0(u)}{u} )\nn \:,
\eeq 
This is the Fourier decomposition of a well-known
{\em Green function}
of the {\em two-dimensional Laplace operator}. This  distributional 
Fourier decomposition works  when the logarithm in the $x_t$ space
acts as an integral kernel 
on a $L^1$ test function $f(x_t)$, provided this  remain
 in $L^1$ when multiplied with the 
logarithm and  have a Fourier transform $\hat f(k_t)$ vanishing at
$k_t =0$ $^{k)}$.
In this case the following integrals exist and it holds:
\begin{eqnarray}
\int_{\R^2} dk_t \: \frac{\hat f(k_t)}{k_t^2} =
 2\pi \int_{\R^2} dx_t \: \ln \frac{|x_t|}{\mu_0} f(x_t)\nn
\end{eqnarray}
In particular $\hat f(0)=0$
 trivially holds when the function $f(x_t)$  is a  
Laplacian
of a function which  decays opportunely as $|x_t|\rightarrow 
+\infty$ $^{l)}$.
Thus, we expect that 
the right  hand 
side of Eq.(\ref{F}) can be written as, 
in the cases of $D_{\tau\tau'}$ $D_{\rho\rho'}$,
 $D_{\rho\tau'}$, $D_{\tau\rho'}$:
\beq
W^{+}_{\mu\mu'}(x,x') 
=\frac{1}{2\pi} \int_{\R^{2}} d x''_{t}\:\: 
\ln \frac{|x''_{t}|}{\mu_0}\:D_{\mu\mu'}
W^{+}(\tau-\tau',\rho,\rho',|x''_{t}-(x_t'-x_t)|)  \label{Fconvolution'}\:,
\eeq
provided the function
$D_{\mu\mu'}
W^{+}(\tau-\tau',\rho,\rho',x''_{t})$ be a 
Laplacian of a function which decays as required.
Notice also that the integrand belongs to $L^1(\R^2)$ as one can verify
by a direct calculation from the asymptotc behaviour of $D_{\mu\mu'}W^+(x_t)$
by Eq.(\ref{W}).\\
Let us consider the action of the operator 
$D_{\tau\tau'}=-\rho\rho'D_{\rho\rho'}$ and thus the explicit
expressions of the functions $W^{+}_{\tau\tau'}(x,x')$ and 
$W^{+}_{\rho\rho'}(x,x')$.\\
In {\bf Appendix B} we shall prove the following remarkable identity:
\beq
D_{\tau\tau'} W^{+}(\tau-\tau',\rho,\rho',x_{t}) = \rho\rho'\: \nabla_{t}^{2}
[\cosh \alpha(\rho,\rho',x_{t}) \:W^{+}(\tau-\tau',\rho,\rho',x_{t})]\:.
\label{fine} 
\eeq
Using the fact that $\ln|x_t|$ is a Green function $^{m)}$ of $\nabla^2_t$,
i.e.
\beq
\frac{1}{2\pi} \int_{\R^2} dx_t \ln |x_t|\: \nabla^2_t g(y_t-x_t) = 
\frac{1}{2\pi} \int_{\R^2} dx_t \ln |y_t-x_t|\: \nabla^2_t g(x_t) =
-g(y_t) \nn
\eeq
and reminding  Eq.(\ref{Fconvolution'}),
it arises:
\beq
W^{+}_{\tau\tau'}(x,x')= -\frac{W^{+}_{\rho\rho'}(x,x')}{\rho\rho'}=
-\rho \rho'\: \cosh \alpha(\rho,\rho',
x_t-x'_t)\: W^{+}(\tau-\tau',\rho,\rho',x_t-x'_t) 
\:. \nn\eeq
Let us consider the action of the operator
$D_{\tau\rho'}$ and thus the explicit
expressions of the functions $W^{+}_{\tau\rho'}(x,x')$ and
$W^{+}_{\rho\tau'}(x,x')$.\\
 It is possible to prove another  remarkable identity,
(see {\bf Appendix B})
namely:
\beq
D_{\tau\rho'} W^{+}(\tau-\tau',\rho,\rho',x_{t}) =
-\rho (\tau-\tau') \nabla^2_t \left(\frac{\sinh \al}{\al} W^+\right)
\label{fine2}
\eeq
and thus, dealing with as in the previously considered case:
\beq
W^{+}_{\tau\rho'}(x,x')= - \frac{\rho W^{+}_{\rho\tau'}(x,x')}{\rho'} =
\rho\:(\tau-\tau')\:\frac{\sinh \alpha
(\rho,\rho',x_t-x'_t)}{\alpha}\: W^+(\tau,\tau',\rho,\rho',x_t-x'_t) 
\:. \nn
\eeq
The cases of $D_{yy'}$ and $D_{zz'}$ are very trivial. In fact, the action of
these operators on the integrand in Eq.(\ref{F})
cancels against the term $k_{\perp}^{-2}$. Thus we have:
\beq
W^+_{yy}(x,x')=W^+_{zz}(x,x')= W^+(x,x') \:.
\nn 
\eeq

Summarizing,
the following Wightman functions calculated on  
Fulling vacuum state arise:
\begin{eqnarray}
W^+_{\tau\tau'}(x,x')
&=& - \rho\rho'\: W^+_{\rho \rho'}(x,x') = \frac{-1}{4\pi^2} 
\frac{\al \coth \al}{\al^2-(\tau-\tau'-i\ep)^2}\label{funo} \:,\\
W^+_{\rho\tau'}(x,x')&=& - \rho'\rho^{-1}\: W^+_{\tau \rho'}(x,x')=
\frac{-1}{4\pi^2\rho}
\frac{\tau-\tau'}{\al^2-(\tau-\tau'-i\ep)^2}\label{fdue}\:,\\ 
W^+_{yy'}(x,x')&=& W^+_{zz'}(x,x') = W^+(x,x') = 
\frac{\al}{4\pi^2 \rho\rho'\sinh\al}
\frac{1}{\al^2-(\tau-\tau'-i\ep)^2} \:.\label{ftre}
\end{eqnarray}
Notice that $
W^{-}_{\mu\mu'}(x,x') = (W^{+}_{\mu\mu'}(x,x'))^{\ast}$.

Let us consider the {\em analytic structure} of the Wightman functions
$W^+$ and $W^-$ extended to the whole {\em complex}
$\tau-\tau'$ plane, 
when $\rho,\rho'\in (0,+\infty) ,x_t,x_t'\in {\R}^2$ are fixed.
We have to consider 
$\ep=0$ and  $\tau-\tau'\rightarrow z = \tau-\tau'+i(s-s')$
($\tau,\tau',s,s'\in \R$) a generally complex number.\\
The structure is the same as in the scalar and 
massless case \cite{full87-152-135}.
It is possible to  extend both the functions 
on the time complex plane except for the possible
appearance of simple poles (instead of branch points)
situated at 
\beq
z = \tau-\tau'= \alpha(\rho,\rho',x_t,x'_t) \:\:\:\:\mbox{namely}
\:\:\:\:
(x-x')^2+|x_t-x'_t|^2-(t-t')^2=0\:. \nn
\eeq
Then, the poles  appear just in case of 
{\em light-like} related arguments. Hence,
the extended
functions $W^+_{\mu\nu'}$ and $W^-_{\mu\nu'}$ result to be  holomorphic
on the whole  remaining complex $z$
plane and both determine the  same analytic continuation on 
just {\em one} Riemann sheet.   In other terms each of these functions is an
analytic continuation of the other.\\
We will term the shared extended function the {\em master function} 
${\cal G}_{\mu\mu'}(\rho,\rho',x_t,x'_t,z)$ \cite{full87-152-135} 
where $z\in \C$. This reads: 
\beq
{\cal G}_{\mu\mu'}(\rho,\rho',x_t,x'_t,z) = V_{\mu\mu'}
(\rho,\rho',x_t,x'_t,z) 
{\cal G}(\rho,\rho',x_t,x'_t,z) \label{master}\:, 
\eeq
where ${\cal G}$ is the master functions, built up dealing with the
same method, of a massless scalar field:
\beq
{\cal G}(\rho,\rho',x_t,x'_t,z) := 
\frac{\al}{4\pi^2 \rho\rho'\sinh\al}
\frac{1}{\al^2-z^2} \:,\nn
\eeq
and the non vanishing bi-vectors $V_{\mu\mu'}$ are:
\beq
V_{\tau\tau'}(\rho,\rho',x_t,x'_t,z) =
- \rho\rho'V_{\rho\rho'}(\rho,\rho',x_t,x'_t,z)  =
 -\rho\rho' \cosh \al\:,\nn
\eeq
\beq
V_{\rho\tau'}(\rho,\rho',x_t,x'_t,z) 
=- \rho'\rho^{-1}V_{\tau\rho'}(\rho,\rho',x_t,x'_t,z)  
= -\rho'(\tau-\tau')
\frac{\sinh\al}{\al}
\:,\nn
\eeq
\beq
V_{yy'}(\rho,\rho',x_t,x'_t,z)  = V_{zz'}(\rho,\rho',x_t,x'_t,z)  = 1
 \:.\nn
\eeq
The functions
$W^+_{\mu\mu'}$ and $W^-_{\mu\mu'}$ are then obtained from
${\cal G}_{\mu\mu'}$ by restricting the complex argument $z$ to the real axis
avoiding the poles from the lower or the  upper $z$ 
complex semiplane  respectively. This approach to the $z$ real axis is
represented by the $\ep-$prescription (also in a distributional sense).

\subsection*{B. PROPAGATOR, SCHWINGER FUNCTION, ADVANCED MINUS RETARDED
FUNCTION }

It is possible to define the photon {\em Feynman propagator} 
$G_{F}(x,x')_{\mu\mu'}$ by 
evaluating ${\cal G}_{\mu\mu'}$
 on the  imaginary $z$ axis
followed by an anticlockwise  rigid rotation of the domain
from the 
imaginary
axis to the real axis \cite{full87-152-135}. Equivalently one can write 
down \cite{full87-152-135}:
\beq
i\:G_{F}(x,x')_{\mu\mu'} := \theta(\tau-\tau') \: W^+_{\mu\nu'}(x,x')
+ \theta(\tau'-\tau) \: W^-_{\mu\nu'}(x,x') \nn\:.
\eeq
Employing  the Klein-Gordon equations which are
satisfied by the  Wightman functions above written,
 remembering that the derivative 
of a  theta function is a  delta function and moreover, using
the  canonical commutation relations Eq.(\ref{CCR}), we obtain also
\beq
g_{\alpha\beta}(x)\nabla^{\alpha}_x \nabla^{\beta}_x G_{F}(x,x')_{\mu\mu'}
= g_{\mu\mu'}(x)\:(-g(x))^{1/2}\:\delta(x,x')\nn\:.
\eeq
This equation holds for test functions with support inside of $W_R$ (considered
as an open set).
Thus, that propagator is a proper {\em Green function} of
vectorial and massless K-G equation as we expected.\\
Finally, let us define the {\em two-point Schwinger function}
as (there is {\em no} summation over  repeated indexes)
\beq
S_{\mu\mu'}(\rho,\rho',x_t,x'_t,s-s') :=
s(\mu) \:s(\mu')\: {\cal G}_{\mu\nu'}(\rho,\rho',x_t,x'_t,i(s-s'))
\nn\:, 
\eeq
where $\rho,\rho' \in (0,+\infty)$, $x_t,x'_t\in \R^2$ and $s,s'\in \R$
and  we defined:
\begin{eqnarray}
s(\sigma) &:= &-i \:\:\:\:\:\mbox{if}\:\:\:\:\: \sigma = 0 \:\:
(\equiv \tau)\nn\\
s(\sigma) &:= &1 \:\:\:\:\:\mbox{if}\:\:\:\:\: \sigma = 1,2,3 \:\:
(\equiv \rho, x_t) \nn\:.
\end{eqnarray}
This Euclidean function is  real and  decays as 
$|s-s'|\rightarrow \infty$. 
We can write $S_{\mu\mu'}(x,x')$ as
\beq
S_{\mu\mu'}(x_E,x_E') = V_{\mu\mu'}(x_E,x_E')\:S(x_E,x_E')\nn
\eeq
where $x_E:=(s,\rho,x_t)$ and the Euclidean bi-vectors 
$V_{\mu\mu'}$ are trivially defined. 
\beq
S(x_E,x_E'):= \frac{\al}{4\pi^2 \rho\rho'\sinh\al}
\frac{1}{(s-s')^2 +\al^2}\nn 
\eeq
is the well-known {\em scalar} Schwinger function in the Rindler
wedge \cite{qualcuno} satisfying:
\beq
g^E_{\alpha\beta}(x_E)
\nabla^{\alpha}_{E} \nabla^{\beta}_{E} S(x_E,x'_E)
= -(g^{E}(x_E))^{-1/2} \:\delta(x_E,x'_E)\:. \nn
\eeq
Starting from the latter equation, some calculations lead us 
to $^{n)}$
\beq
g^E_{\alpha\beta}(x_E)
\nabla^{\alpha}_{E} \nabla^{\beta}_{E} S_{\mu\mu'}(x_E,x'_E)
= -g^E_{\mu\mu'}(x_E)\:(g^{E}(x_E))^{-1/2}
\:\delta(x_E,x'_E)\:,\nn
\eeq
where $g^{E}_{\mu\nu}:=$ diag$(+\rho^2,1,1,1)$ is the
{\em Euclidean metric} associated with the initial Rindler metric and 
the covariant derivative is defined with respect to this metric
$g^{E}_{\mu\nu}$.  Thus $S_{\mu\mu'}$ is an {\em Euclidean Green function}
decaying as 
$|s-s'|\rightarrow \infty$
of the vectorial K-G equation on test functions with support in
$\{\rho\in (0,+\infty), x_t\in {\R}^2, s\in \R\}$. 
Note  the points with $\rho=0$ are singular points 
of the Euclidean manifold.
We have defined our Euclidean manifold in order to exclude 
these points. 
All that we have found  is very similar to
 the case of a scalar field propagating
in the whole Minkowski manifold as well as in the Rindler Wedge 
 \cite{fulling,full87-152-135,birrelldavies}.\\

Finally, let us consider the {\em advanced minus retarded} fundamental solution
 namely, the fields operators  commutator. We shall deal with
 in contravariant components.
\beq
E^{\mu\mu'}(x,x') := W^{+\mu\mu'}(x,x') - W^{-\mu\mu'}(x,x')
\:\:\:= [\hat{A}^\mu (x), \hat{A}^{\mu'}(x')] 
\label{E}\:. 
\eeq 
We have from Eq.s (\ref{funo}), (\ref{fdue}), (\ref{ftre}) $^{o)}$:
\begin{eqnarray}
E^{\tau\tau'}(x,x')&=& -\frac{1}{\rho\rho'}\:E^{\rho\rho'}(x,x')\:\:=\:\:
\frac{i\:\alpha\:\coth \alpha}{2\pi\:\rho^2\rho'^2}
\:\mbox{sign}(\tau-\tau')\: \delta(\alpha^2-(\tau-\tau')^2) \:, \label{Euno}\\
E^{\rho\tau'}(x,x')&=&-\frac{\rho}{\rho'}\:E^{\tau\rho'}(x,x')\:\:=\:\:
\frac{i(\tau-\tau')}{2\pi\: \rho\rho'^2}
\:\mbox{sign}(\tau-\tau')\: \delta(\alpha^2-(\tau-\tau')^2) \:, \label{Edue}\\
E^{yy'}(x,x') &=& E^{zz'}(x,x')\:\:=\:\:
\frac{-i\alpha}{2\pi\:\rho\rho'\:\sinh\alpha}
\:\mbox{sign}(\tau-\tau')\: \delta(\alpha^2-(\tau-\tau')^2) \:, \label{Etre}
\end{eqnarray}
It is possible to prove that the  vectorial
advanced minus retarded fundamental solution reduces to the
 Minkowskian one, when  the domain of test functions is
restricted to the Rindler Wedge.
We shall just sketch a proof of this in the following. \\
The advanced minus retarded 
solution of a
photon field 
propagating in the
 whole Minkowski space results to be
 written $^{p)}$
in our initial  Minkowskian coordinates $(x^0,x^1,x^2,x^3) 
\equiv (t,x,x_t) \equiv (t,\vec{x})$
as
\beq
E_M^{\mu\nu}(x,x') = \eta^{\mu\nu} \:\frac{-i}{2\pi}\: 
\mbox{sign}(t-t')\: \delta((t-t')^2-|\vec{x}-\vec{x}'|^2) \nn \:.
\eeq
Remaining in a Minkowskian base of the tangent space,
 but passing to Rindler coordinates as far as the arguments
of the functions are concerned, 
employing standard distributional manipulations,
one can also write down $E_M^{\mu\nu}(x,x')$
in $W_R$ as $^{q)}$
\beq
E_M^{\mu\nu}(x,x') = \eta^{\mu\nu} \:\frac{-i\:\alpha }{2\pi\:\rho\rho'\:
\sinh\alpha}\:
\mbox{sign}(\tau-\tau')\: \delta(\alpha^2-(\tau-\tau')^2)
\label{Equattro'} \:,
\eeq 
Starting from $E^{\mu\mu'}$ expressed in Rindler coordinates by 
Eq.s (\ref{Euno}), (\ref{Edue}) and (\ref{Etre}) and 
coming back to Minkowski tetrad, we find just the right hand side of 
Eq.(\ref{Equattro'}).
Take into account that, because of the presence of
a {\em delta} function in Eq.s (\ref{Euno}), (\ref{Edue}) and (\ref{Etre}),
it is possible to change $|\tau-\tau'|$ with $\alpha$ (and so on)
during  calculations. 

\subsection*{C. A WARD IDENTITY}

Let us consider the identity (where primed derivatives works on
primed arguments):
\beq
g_{\mu \sigma}(x)
\nabla^{\sigma} G_F^{\mu\nu'}(x,x') - \nabla^{\nu'} G_F(x,x') =0 
\label{ward}\:,
\eeq
where $G_F$ is the scalar massless Feynman  propagator.
Reminding that equal time evaluated fields operators commute and
the definition of Feynman propagator in therms of Wightman functions,
the  formula written above
results to be equivalent to:
\beq
\nabla^{\mu} W^\pm_{\mu\nu'}(x,x') = -g_{\nu'\lambda'}(x')
\nabla^{\lambda'} W^\pm(x,x')  \label{finesection}\:.
\eeq
The identity in Eq.(\ref{ward}) is very important because it is a
{\em Ward identity}  for the photon field in the Feynman gauge
obtained (in Minkowskian coordinates) by a path integral quantization and 
imposing the BRST 
invariance \cite{libroverde} $^{r)}$.\\
It is possible to prove Eq.(\ref{finesection}) by explicitly calculating
both sides through the  formulae obtained above. This proof does not contain
intersting comments and we do not report on this here. Conversely, we shall
report a less rigorous  but physically more interesting
  proof of Eq.(\ref{finesection}). This ``proof'' points out
the role of physical and unphysical modes in  Ward's identity.  \\
Holding Eq.(\ref{za}), it is necessary to prove only that
(the proof for the case of $W^-$ is identical):
\beq
\nabla^\mu \int \: d\omega \:dk_t [ A^{(3,\omega,k_t)}_{\mu}(x)
A^{\ast(3,\omega,k_t)}_{\mu'}(x') 
-A^{(0,\omega,k_t)}_{\mu}(x)A^{\ast(0,\omega,k_t)}_{\mu'}(x')] 
= - \partial_{\mu'}\: W^{+}(x,x') \nn \:.
\eeq
Employing the modes $A^G$ and $A^L$ which appear in Eq.s (\ref{ezero}) and
(\ref{etre}), 
and noticing that $\nabla^\mu A^G_\mu=0$, 
the identity above written
reduces to:
\beq
i \int \: d\omega \:dk_t \nabla^\mu A^{(L,\omega,k_t)}_{\mu}(x)
A^{\ast(G,\omega,k_t)}_{\mu'}(x') = - \partial_{\mu'}
\: W^{+}(x,x') \nn \:.
\eeq
Expanding the covariant derivative in the  integrand  and evaluating
the modes $A^G$ and $A^L$ in terms of the field $\phi$ by Eq.s (\ref{atre})
and (\ref{aquattro}), 
the identity 
to be proved reads
\beq
- \int \: d\omega \:dk_t \frac{\sinh \pi\omega}{4\pi^4}\:
\phi^{(\omega,k_t)}(x) \partial_{\mu'} 
\phi^{\ast (\omega,k_t)}(x') = - \partial_{\mu'}\: W^{+}(x,x') \nn \:.
\eeq 
This holds by definition of $W^+$.


\section*{IV. THERMAL GREEN FUNCTIONS AND SUBTLETIES WITH GAUGE INVARIANCE}
 
\subsection*{A. PHOTON KMS STATES}

Dealing with static coordinates $(\vec{x},t)$ in a  spatially finite
static region of a space time, the scalar
{\em thermal} Wightman functions are defined 
as: 
 (see for example \cite{birrelldavies})
\beq
W^+_\be(\vec{x},\vec{x'}, t-t') = Z^{-1}_\be 
 Tr \{ e^{-\be \hat H} \hat\phi (\vec{x},t) \hat\phi (\vec{x}',t')\}\:,	
\label{wwa}
\eeq
\beq
W^-_\be(\vec{x},\vec{x'}, t-t') = Z^{-1}_\be
 Tr \{ e^{-\be \hat H} \hat\phi (\vec{x'},t')\hat\phi (\vec{x},t)\}\:,
\label{wwb}
\eeq
where $Z_\be := e^{-\be\hat H}$ is the partition function of the field at
temperature $^{s)}$ $T_0= 1/\be$.\\
These 
 formulae 
have to be opportunely  mathematically interpreted due to 
``operator'' $\hat\phi\hat\phi$ 
which is not a (trace class) bounded operator. However we shall not discuss
 on this
here, because our discussion has to be understood just in an
heuristic sense (for details see \cite{full87-152-135,haaglibro}
 and Ref.s therein).
By extending the thermal Wightman functions  defined above
 to the complex time, we can recover the
{\em KMS condition} \cite{KMS}
 due to cyclic property of the trace \cite{full87-152-135}
 \cite{birrelldavies}:
\beq
W_\be^{\pm}(\vec{x},\vec{x'}, t-t'\mp i\be) =
 W^{\mp}_\be(\vec{x},\vec{x'}, t-t')
\nn \:.
\eeq
Provided appropriate mathematical conditions hold 
\cite{full87-152-135}, these  Wightman 
functions can be continued into an analytic function, the {\em thermal} 
master function
${\cal G}_\be(\vec{x},\vec{x}',z)$, defined in the time complex plane
$z = t-t'+ i(s-s')$, 
periodic in the imaginary time  $s-s'$ with period $\be$.
This function results to be defined on the whole $z$
plane except for cuts on the real axis (periodically repeated along the
imaginary axis, see figures in \cite{full87-152-135}) corresponding
to light-related arguments. The cuts terminate on branch points which become
simple poles in the case of a massless field. The Wightman functions $W_\be^+$
and $W_\be^-$
result to be defined by approaching the real axis respectively
from the lower semiplane and the upper semiplane (following the
 $\ep-$prescription). The discontinuity crossing the cuts
 gives rise  to the coincidence of 
the difference of the two Wightman functions  
and the ($\be$ independent) advanced minus retarded fundamental solution.\\  
In case of fields propagating inside of an {\em infinite} spatial volume the
partition function defined as a trace does not exist. However  other
possible definitions 
follow from path integral (and $\zeta$ 
function or heat-kernel methods) but this is not our case.
 Following
\cite{KMS} (see also Ref.s \cite{kaywald,haaglibro})
  the (quasifree) scalar 
thermal
states can be defined, by an {\em algebraic} approach in terms of $*-$, Weyl,
$C^*-$
 and Von Neuman algebras as functionals on the algebra of the field. 
In this case, provided appropriate mathematical requirements be satisfied 
\cite{full87-152-135}, the thermal Wightman functions
are (positive) 
integral kernels bi-solutions of the motion equations which satisfy
 the KMS condition written above,
having the analytic structure previously pointed out and producing the
advanced minus retarded fundamental solution by difference.
  Hence, one can use
the integral kernels to built up the (quasifree) state as a positive
 functional on the
($*-,$ etc.) algebra generated by the field.\\ 
The algebraic way to define thermal
Wightman functions and thermal states agrees with the naive
procedure (based 
on Eqs.(\ref{wwa}) and (\ref{wwb})) whenever that can be 
implemented in some sense.
In particular, when the naive method is  correctly used in a finite box
with convenient boundary conditions and the box walls are moved away 
to infinity in the end
\cite{full87-152-135,haaglibro}.

Other remarkable facts are also important. 
It is possible to prove that  
 the thermal master function
${\cal G}_\be(\vec{x},\vec{x}',z)$
 evaluated on the imaginary time axis, the Schwinger function  
${\cal S}_\be(\vec{x},\vec{x}',s):= $
${\cal G}_\be(\vec{x},\vec{x}',is)$, coincides with 
an imaginary time {\em periodic} Green function of the
Euclidean Laplace operator. This  operator is defined  in the imaginary time 
periodic
Euclidean section 
of the manifold with
 period $\beta$. In this way,
the previously written KMS conditions directly follow from the imaginary time
 periodicity of the manifold. \\
Another important point is the {\em sum over images method}. It is well-known
\cite{full87-152-135} that the above considered extended 
periodic Green functions can be obtained from
the non thermal ones by the sum:
\beq
{\cal G}_\be(\vec{x},\vec{x}', z) =
\sum_{n \in \Z} {\cal G}(\vec{x},\vec{x}', z+ i n\be) 
\nn\:,
\eeq
where ${\cal G}(z)$ is the analytic extension to the complex time of
 the {\em non} thermal master function. 
Furthermore it can be proved \cite{full87-152-135}
${\cal G}_{\infty}(z) = {\cal G}(z)$,
 where $\infty$ denotes the limit as $\beta\rightarrow 
+\infty$.
All these topics have been more or less rigorously 
implemented in the scalar  and
spinorial case
  in different manifolds and, in particular, in the Rindler wedge
for massless fields, also in relation to the  cosmic string theory
(see e.g.  Ref.s \cite{robavanzo,full87-152-135,mova,linet} 
and ref.s therein).\\

Let us consider the case of a photon field in  Feynman's
 gauge.\\
Obviously, it is possible to directly 
define strength field $F_{\mu\nu}$ Wightman functions
  avoiding unphysical particles and 
gauge related problems. However,
this is not a completely satisfactory way because, for instance,
implementing an interaction
theory one must use directly the field $A_\mu$ in dealing with
the minimal coupling.\\
We shall start by supposing to work within a finite box in order to have a
well defined partition function and to be able to use the naive formalism.
The following discussion is just heuristic, no mathematical rigor
is used.\\
  The hardest problem is due to the presence of unphysical degrees of freedom.
Such a difficulty has been pointed out by Bernard \cite{bernard}
dealing with the Euclidean path integral formalism to define the photon
partition functions in an arbitrary gauge. He proved that
the correct definition, not depending on the gauge, is the trace over the
 physical degrees of freedom only: 
\beq
Z_{\mbox{phys}\:\be} =
 \sum_{\Psi_n \:\:\mbox{phys.}} <\Psi_n| e^{-\be \hat H} |\Psi_n>
\nn \eeq 
Successively, 
this definition has to be re-written
as a path integral in the chosen gauge by a Faddeev-Popov ghost 
procedure 
in such
a manner to include the unphysical modes in the functional integral.
Following this way, we can start by formally define in a large box in 
the Rindler wedge:
\beq
W^{\mbox{phys.}\:+}_{\be\mu\mu'}(\rho,\rho',x_t,x_t', \tau-\tau')
 :=  \nn
\eeq
\beq
= Z_{\mbox{phys}\:\be}^{-1}
\sum_{\Psi_n \:\:\mbox{phys.}}<\Psi_n| e^{-\be \hat H} 
\hat A^{\mbox{phys.}}_{\mu}(\rho,x_t,t) \hat A^{\mbox{phys.}}_{\mu'}
(\rho',x_t',t')|\Psi_n>\:,
\label{www1}\eeq
and 
\beq
W^{\mbox{phys.}\:-}_{\be\mu\mu'}(\rho,\rho',x_t,x_t', \tau-\tau')
 := \nn
\eeq
\beq
= Z_{\mbox{phys}\:\be}^{-1}
\sum_{\Psi_n \:\:\mbox{phys.}}<\Psi_n| e^{-\be \hat H}
\hat A^{\mbox{phys.}}_{\mu'}(\rho',x_t',t') \hat A^{\mbox{phys.}}_{\mu}
(\rho,x_t,t)|\Psi_n>\:,
\label{www2}\eeq
where $\hat A^{\mbox{phys.}}_\mu$ contains only the transverse 
modes, i.e., $\lambda =1$ 
and $2$ and $|\Psi_n>$ denotes the eigenvector of $\hat H$  with eigenvalue
$E_n$.\\
Now we can add to these Wightman functions 
an unphysical part related to the
considered gauge choice, Feynman gauge in the present case.
 This part has to vanish when the Wightman functions act on physical 
 wavefunctions.  Such a procedure must not affect the Wightman
functions calculated by the strength field $F_{\mu\nu}$.\\  
Our proposal consists of the  formal definition
 (the definition of $W^-$ being obvious):
\beq
W^{\mbox{Feynman}+}_{\be\mu\mu'}(\rho,\rho',x_t,x_t', \tau-\tau')
 := \nn
\eeq
\beq
 = Z_{E\be}^{-1}
\sum_{n}<\Psi_n/ e^{-\be \hat H}
\hat A_{\mu}(\rho,x_t,t) \hat A_{\mu'}(\rho',x_t',t')/\Psi_n>\:,
\label{ourdef}
\eeq
where 
\beq
Z_{E\be} = Tr_E e^{-\be \hat H} := \sum_n <\Psi_n/ e^{-\be \hat H}/\Psi_n>\:,
\nn \eeq
the index $E$ denotes the use of the Euclidean scalar product in calculating
the trace above.
 Notice that $\hat H$ is non negative employing the Euclidean scalar product,
in fact we have:
\beq
\hat H =  \int dk_t \: d\omega \: \omega \:
\sum_{\lambda=0}^3 \eta^{\lambda\lambda}\:
\hat a^{\dagger}_{(\lambda,\omega,k_{t})} \hat a_{(\lambda,\omega,k_{t})}
= \int dk_t \: d\omega \: \omega \:
\sum_{\lambda=0}^3 \delta^{\lambda\lambda}\:
\hat a^{\star}_{(\lambda,\omega,k_{t})} \hat a_{(\lambda,\omega,k_{t})}
\:,\nn
\eeq
and thus
no  problem on the divergence of the trace arises.
It is quite simply proved that formally:
\beq
W^{\mbox{Feynman}+}_{\be\mu\mu'} = 
W^{\mbox{phys.}\:+}_{\be\mu\mu'}
+ Z^{-1}_{\mbox{unph.}}
\sum_{\Psi_n \:\:\mbox{unph.}}<\Psi_n/ e^{-\be \hat H}
\hat A^{\mbox{unph.}}_{\mu} \hat A^{\mbox{unph.}}_{\mu'}/\Psi_n>\:,
\nn \eeq
where the  vacuum state is included in the sum over
unphysical states.
Notice that the second term vanishes employing physical test wavefunctions.
Using such wavefunctions the Wightman functions are also positive
defined by construction. We also stress that, by the definition
Eq.(\ref{ourdef})
the difference of the two Wightman functions 
does not depend on $\be$ and reproduces the non thermal commutator.
Finally, the thermal strength Wightman functions
calculated as  derivative of 
 the Wightman functions defined in Eq.(\ref{www1})
and (\ref{www2})
coincide with those obtained by the derivatives of 
the physical Wightman functions
defined in Eq.(\ref{ourdef}). This is just a trivial consequence of
$F^{(G)}_{\mu\nu}(x)=F^{(L)}_{\mu\nu}(x)=0$.\\
Our definition trivially satisfies K-G equations 
and  maintains the KMS condition due to cyclic property
of the trace  involved in Eq.(\ref{ourdef}). 
Following the way employed in \cite{full87-152-135}, we expect 
to find also the analytic structure previously pointed out. 
Furthermore, the Ward identity
in Eq.(\ref{ward}) can be formally proved employing the same method.
 Finally, one 
finds the non thermal Wightman functions as the limit
$\beta \rightarrow +\infty$.\\
We stress that different proposals of definition involving, in 
Eq.(\ref{ourdef}),
the physical
scalar product defined in Eq.(\ref{scalar2}), 
instead of the Euclidean one, do not
maintain the KMS condition.
This is due to the presence of the
operator $M$ which does not permit to take advantage of the cyclic property
of the trace.

\subsection*{B. THERMAL WIGHTMAN FUNCTIONS AND RELATED THERMAL GREEN FUNCTIONS}

Taking  account of the  heuristic discussion 
performed above, we shall define 
the thermal Wightman functions of the photons in the Feynman gauge
by requiring they are bi-solutions of K-G equations, 
satisfy the KMS condition, take on the analytic
structure of the scalar Wightman functions and produce the
advanced minus retarded fundamental solution of Eq.(\ref{E}) by the usual
difference. \\
We shall try to built up such Wightman functions by a thermal master
function obtained by summing over 
images.
Remind the series:
\beq
\sum_{n\in \Z} \frac{1}{(a+n)^2-b^2} = \frac{\pi}{2b} 
\left\{ \cot \left[\pi(a-b)\right]-\cot\left[\pi(a+b)\right]\right\} \nn\:,
\eeq
absolutely convergent, for $a, b \in\C$ such that both sides are defined
and 
\beq
\sum_{n\in \Z} \frac{a+n}{(a+n)^2-b^2} = \frac{\pi}{2} \left\{\cot
\left[\pi(a-b)\right] +
\cot\left[\pi(a+b)\right]\right\} \nn\:,
\eeq
convergent 
in the sense of the principal value (namely, $\lim_{N\rightarrow +\infty}
\sum_{|n|<N}$),
 for $a, b \in\C$ such that both sides are defined.
Then, let us consider the thermal master function defined 
as
\beq
{\cal G}_{\be \mu\mu'}(\rho, \rho',x_t,x'_t, z) := \sum_{n\in \Z}
{\cal G}_{\mu\mu'}(\rho, \rho',x_t,x'_t, z+in\be) 
\:, \label{mastert} 
\eeq
where ${\cal G}_{\mu\mu'}(\rho, \rho',x_t,x_t, z)$ was defined 
in Eq.(\ref{master}). The convergence is understood as punctual convergence
 in the 
sense of the principal value at least. \\
Employing the  results above as well as trivial  calculations
we find (re-arranging the result in a convenient form for future reference):
\beq
{\cal G}_{\be\: \tau\tau'}(z)=\frac{-1}{4\pi\be \sinh\al}
\frac{\cosh\left(\frac{2\pi}{\be}
z\right)\sinh\al+\sinh\left[\left(\frac{2\pi}
{\be}-1\right)\al\right]}{\cosh\left(\frac{2\pi}{\be}\al\right)-
\cosh\left(\frac{2\pi}{\be}z\right)} 
- \frac{1}{4\pi \be} \label{mastert1}\:,
\eeq
\beq
{\cal G}_{\be\:\rho\rho'}(z)= -\frac{1}{\rho\rho'}{\cal G}_{\be\: \tau\tau'}(z)
\label{mastert2}\:,
\eeq
\beq
{\cal G}_{\be\: \rho\tau'}(z) =
-\frac{1}{4\pi\be\rho}\frac{\sinh\left(\frac{2\pi}{\be}
z\right)}{\cosh\left(\frac{2\pi}{\be}\al\right)-
\cosh\left(\frac{2\pi}{\be}z\right)}
\label{mastert3}\:,
\eeq
\beq
{\cal G}_{\be\: \tau\rho'}(z) = -\frac{\rho'}{\rho}
{\cal G}_{\be\: \rho\tau'}(z)
\label{mastert4}\:,
\eeq
\beq
{\cal G}_{\be\: yy'}(z) ={\cal G}_{\be\: zz'}(z) ={\cal G}_{\be}(z) 
\label{mastert5}\:.
\eeq
${\cal G}_{\be}(z)$ is the thermal master function of a massless scalar
field obtained summing over images the previously calculated 
non thermal master function.
Notice the periodicity $\be$ in the imaginary time.
We can consider  $z= \tau-\tau' \pm i\ep$ to obtain the 
thermal Wightman functions. We report $W^+_{\be\:\mu\mu'}$ only,
 $W^-_{\be\:\mu\mu'}$ is obtained
by a complex conjugation of the former.
\beq
W^{+}_{\be\:\tau\tau'}(x,x')
=\frac{-1}{4\pi\be\sinh\al}
\frac{\cosh\left(\frac{2\pi}{\be}
(\tau-\tau^{'})\right)\sinh\al+\sinh\left[\left(\frac{2\pi}
{\be}-1\right)\al\right]}{\cosh\left(\frac{2\pi}{\be}\al\right)-
\cosh\left(\frac{2\pi}{\be}(\tau-\tau^{'}-i\ep)\right)} 
- \frac{1}{4\pi \be} \label{wt1}\:,
\eeq
\beq
W^{+}_{\be\:\rho\rho}(x,x')=-\frac{1}{\rho\rho'}W^{+}_{\be\:\tau\tau'}(x,x')
\label{wt2}\:,
\eeq
\beq
W^{+}_{\be\:\tau\rho'}(x,x')=
-\frac{1}{4\pi\be\rho'}\frac{\sinh\left(\frac{2\pi}{\be}
(\tau-\tau^{'})\right)}{\cosh\left(\frac{2\pi}{\be}\al\right)-
\cosh\left(\frac{2\pi}{\be}(\tau-\tau'-i\ep)\right)}
\:, \label{wt3}
\eeq
\beq
W^{+}_{\be\:\rho\tau'}(x,x') = -\frac{\rho'}{\rho}
W^{+}_{\be\:\tau\rho'}(x,x')
\:, \label{wt4}
\eeq
\beq
W^{+}_{\be\:yy'}(x,x')= W^{+}_{\be\:zz'}(x,x')=W^{+}_{\be}(x,x')
\label{wt5}\:.
\eeq
 $W^{+}_{\be}(x,x')$ is the well-known Rindler thermal
Wightman function of a massless scalar field \cite{mova,qualcuno}:
\beq
W^{+}_{\be}(x,x') = 
\frac{1}{4\pi \be \rho\rho'\sinh \al}\left[ \frac{\sinh \left(\frac{2\pi}{\be}
\al\right)}
{\cosh\left(\frac{2\pi}{\be}\al\right) -\cosh\left(\frac{2\pi}{\be}
(\tau-\tau'-i\ep)\right)}\right]
\nn\:.
\eeq 
The vectorial 
Wightman functions  written above trivially satisfy the KMS condition
and the related thermal master function has the required 
analytic structure. In particular, there are not branch points but  
a pair of  
simple poles periodically repeated in the imaginary time with
period $\be$. The two poles on the real time axis correspond
light-like related arguments $^{t)}$.\\
Moreover, some calculations
 involving standard distributional properties
  prove the difference\\
 $W^+_{\be\: \mu\mu'}(x,x')~-~W^-_{\be\:\mu\mu'}(x,x')$ coincides
with  the advanced minus retarded solution defined in Eq.(\ref{E}). 
We might prove that the obtained vectorial
 Wightman functions define a {\em positive}
bi-functional working on physical  test wavefunctions. We shall prove this
in the case $\be=2\pi$ only.

Some comments on the obtained  functions are necessary.
First let us evaluate the thermal master function along the
time imaginary axis. We obtain the thermal Schwinger function. 
($(x_E^0,x_E^1,x_E^2,x_E^3) \equiv (s, \rho, y,z)$)
\beq
S_{\be\:00'}(x_E,x'_E)
=\frac{1}{4\pi\be\sinh\al}
\frac{\cos\left(\frac{2\pi}{\be}
(s-s')\right)\sinh\al+\sinh\left[\left(\frac{2\pi}
{\be}-1\right)\al\right]}{\cosh\left(\frac{2\pi}{\be}\al\right)-
\cos\left(\frac{2\pi}{\be}(s-s')\right)}
+ \frac{1}{4\pi \be} \label{st1}\:,
\eeq
\beq
S_{\be\:11'}(x_E,x_E')= \frac{1}{\rho\rho'}S_{\be\:00'}(x_E,x_E')
\label{st2}\:,
\eeq
\beq
S_{\be\:01'}(x_E,x_E')=
\frac{-1}{4\pi\be\rho'}\frac{\sin\left(\frac{2\pi}{\be}
(s-s')\right)}{\cosh\left(\frac{2\pi}{\be}\al\right)-
\cos\left(\frac{2\pi}{\be}(s-s')\right)}
\:, \label{st3}
\eeq
\beq
S_{\be\:10'}(x_E,x_E') = -\frac{\rho'}{\rho}
S_{\be\:01'}(x_E,x_E')
\:, \label{st4}
\eeq
\beq
S_{\be\:yy'}(x_E,x_E')= S_{\be\:zz'}(x_E,x_E')=S_{\be}(x_E,x_E')
\label{st5}\:.
\eeq
This bi-vectorial function trivially 
defines a periodic  vectorial Laplacian Green function in
 the Euclidean section of the manifold with imaginary time period $\be$.
This follows from Eq.(\ref{mastert}) when one supposes
 $z=is$, considers the series
as a series of distributions and reminds that the non thermal Schwinger
function is a Green function in the non periodic Euclidean manifold.
The function 
\beq
S_{\be}(x_E,x_E') =
\frac{1}{4\pi \be \rho\rho'\sinh \al}\left[ \frac{\sinh \left(\frac{2\pi}{\be}
\al\right)}
{\cosh\left(\frac{2\pi}{\be}\al\right) -\cos\left(\frac{2\pi}{\be}
(s-s')\right)}\right]
\nn
\eeq
is a Rindler thermal Schwinger function for a massless scalar field
(see also \cite{qualcuno} where a different nomenclature is used)
obtained, for instance, by the  sum over images method.  Few words on this 
function in relation to the vectorial found ones are necessary.
The corresponding
 test functions of the scalar Scwinger function 
have support  in 
$\{ s\in [0 ,\be ),\rho \in [0,+\infty), y,z \in \R^2\} $
(where $0\equiv \beta$).
Differently from the case $\be=+\infty$ (i.e. the non thermal case),
the scalar thermal Schwinger function
 is defined also when $\rho' \rightarrow 0$ and 
$\rho > 0$ (and {\em vice versa}), namely, when one of the arguments stays
 on the
{\em tip} of the Euclidean Rindler cone. Remind that the
 Euclidean Rindler manifold is  diffeomorfic
to ${\cal C}_\be {\bf \times} \R^2$ where the first factor is a cone of
angular deficit $2\pi-\be$. There, $s$ is the angular variable and $\rho$ the
radial one. We have: 
\beq
S_0(x_E, x_t'):= S_\be(x_E,x_E')|_{\rho'\rightarrow 0} 
 =\frac{1}{2\pi \be (\rho^2+|x_t-x_t'|^2)}\:,
\nn\eeq
and  $\nabla^2_E S_0(x_E, x_t') = 0$ whenever $\rho >0$.\\
Employing carefully  second Green's identity, one can quite simply 
prove $^{u)}$: 
\beq
\int_{{\cal C}_\be {\bf \times} \R^2} d^4 x_E \sqrt{g_E(x_E)}\: S_0(x_E,x_t')
\: \nabla_E^2 f(x_E)  = - f(s,\rho', x_t')|_{\rho'=0} 
 \nn\:.
\eeq 
Thus, we see that, in the massless scalar case,  the Schwinger function is a 
Green function for the {\em whole} Euclidean manifold whenever 
$0<\be<+\infty$.   The 
case of the vectorial field is quite different.\\
In order to study that case it is convenient  to write the vectorial
 Schwinger functions in a unitary 
normalized base of the cotangent space (a tetrad).
 This avoids  troubles related to an anomalous normalization of coordinate
 base vectors
in the limit $\rho\rightarrow 0$. 
This vectorial Schwinger function,
by normalizing the base $ds, d\rho, dx_t$, takes on a factor $\rho^{-1}$ 
($\rho'^{-1}$) for each $0-$ ($0'-$) component. The  $11'-$ component,
and the transverse ones  $yy'$, $zz'$
do not change. Except for the case $\be=2\pi$ which we shall study later,
the limit as $\rho'\rightarrow 0$ ($\rho$ fixed) produces vanishing
or infinite non transverse components of the vectorial Schwinger function,
 depending on the sign of $\be-2\pi$.\\
Such an anomalous 
behaviour for the vectorial case 
 seems related to the presence
of the conical singularity on the tip of a cone,
 which does not permit to unambiguously  define
the tangent (cotangent) space and the metric tools there.\\
Let us  directly consider the found  vectorial thermal Wightman functions.
We notice that these reduce to the correct non thermal limit Eq.s (\ref{funo}) 
(\ref{fdue}) and (\ref{ftre}) in the case
$\be \rightarrow +\infty$. Furthermore, 
one can prove by direct calculations that both K-G equations and 
the   Ward identity holding in
the non thermal case Eq.(\ref{finesection}) are
 satisfied.
We do not further report on this here because the proof
does not involve interesting comments.  

\subsection*{C. SUBTLETIES WITH GAUGE INVARIANCE}

Let us consider the strange {\em static} term 
$\delta W_{\beta\:\mu\mu'}(x,x')$ added
to the $\tau\tau'$ and $\rho\rho'$
transversal thermal Wightman functions  (see Eq.s (\ref{wt1}) and 
(\ref{wt2})). In components it reads:
\beq 
\delta W_{\be \:\tau\tau'}(x,x') = -\frac{1}{4\pi\be}\:,\:\:\:\:
\delta W_{\be \:\rho\rho'}(x,x') = \frac{1}{4\pi\be\rho\rho'}
\label{no}
\eeq
(all the remaining components vanish).\\
This term does not contribute to the zero temperature limit because this
vanishes as $\be\rightarrow +\infty$.
Furthermore, this term  is responsible for an apparently
 bad behaviour of the thermal Wightman 
functions as $|x_t-x_t'|\rightarrow +\infty$ when $\be<+\infty$.
In fact, the thermal Wightman functions do not vanish in this limit.
However, considering thermal states, the requirement of a vanishing 
large distance
fields correlation
  is not so strictly necessary. Anyhow, we shall see that
in the present case the {\em physical} correlations do vanish in the 
considered limit, because the terms in Eq.(\ref{no}) do not contribute
to the ``physical part'' of Wightman functions. \\ 
Also notice that, because of the form of
$\delta W_{\beta\:\mu\mu'}(x,x')$, this term does not affect the Wightman
functions calculated by  
the strength field operator $\hat F_{\mu\nu}(x)$. In fact, the contribution
to the strength field thermal Wightman functions 
reads
\beq
\delta < \hat F_{\mu\nu}(x)\hat F_{\mu'\nu'}(x')>_\be=\nn
\eeq
\beq
\nabla_\mu\nabla'_{\nu'} \delta W_{\be \:\mu\mu'}(x,x') -
\nabla_\nu\nabla'_{\mu'} \delta W_{\be \:\mu\nu'}(x,x') -
\nabla_\mu\nabla'_{\nu'} \delta W_{\be \:\nu\mu'}(x,x')
-\nabla_\mu\nabla'_{\mu'} \delta W_{\be \:\nu\nu'}(x,x')\nn
\eeq  
In order to obtain some non vanishing term in this sum,
it must be $\mu=\nu=\rho$  and $\mu'=\nu'=\rho'$. In such a situation
the four terms cancels each others and the final result vanishes.\\
Let us prove $\delta W_{\beta\:\mu\mu'}(x,x')$ contains  
 {\em gauge terms} only, has a vanishing covariant divergence and 
satisfies vectorial Klein-Gordon equations.\\
In particular, the vanishing covariant divergence implies that
 the term 
$\delta W_{\beta\:\mu\mu'}(x,x')$ can be omitted (or that 
we can use
a different value $\be'\neq \be$) in checking the
previously discussed Ward identity.\\
We can write 
\beq
\delta W_{\beta\:\mu\mu'}(x,x') = \nabla_\mu \nabla'_{\mu'} \Phi(x,x')
\:\:\:\:
\mbox{where}\:\:\:\: \Phi(x,x') := \frac{-1}{4\pi\be}(\tau\tau'-\ln \rho
 \ln\rho')\:. \label{deltabeta}
\eeq
Hence, only gauge terms appear in $\delta W_{\beta\:\mu\mu'}(x,x')$.\\  
$\delta W_{\beta\:\mu\mu'}(x,x')$
has a vanishing covariant divergence
because
$\nabla^\mu\nabla_\mu \Phi(x,x') = 0$ due to  Eq.(\ref{deltabeta}).
Furthermore, due the commutativity of covariant derivative inside
of a flat manifold, we find also
$\nabla_\sigma\nabla^\sigma \delta W_{\beta\:\mu\mu'}(x,x') =0$.\\
Finally, let us prove  that the considered term produces no
contribution to the value
of thermal Wightman functions when they act on, {\em at least one},
 physical test
wavefunction. More generally, we shall prove:
\beq
\int_\Sigma dS\: n^\mu \sqrt{h} \: A_\nu(x) \lrup{\nabla}_\mu
\delta W_{\beta}^{\nu\nu'}(x,x') = 0 \label{nocontribute}\:,
\eeq
where $A\in {\cal S}$ satisfies also $\nabla_\mu A^\mu =0$.
This includes the wavefunctions built up employing physical modes
$A^{(1)}, A^{(2)}$
as well as the gauge modes $A^{(G)}$ (see Eq.s(\ref{ezero}), (\ref{euno}),
 (\ref{edue})
and (\ref{etre})), namely {\em physical} and
{\em Lorentz} wavefunctions. An analog proof can be produced out by employing
 a four smeared formalism introduced in {\bf Appendix C} and, in that
 case,
the constraint $\lap_\mu A^\mu=0$ becomes $\lap_\mu F^\mu=0$ where
$F^\mu(x)$ is a four smeared test function.\\
The left hand side of Eq.(\ref{nocontribute}) can be written, due 
to Eq.(\ref{deltabeta}) (omitting the unimportant second 
argument of $\Phi$ and its covariant derivative):
\beq
\int_\Sigma dS\: n^\mu \sqrt{h} \: A_\nu(x) \lrup{\nabla}_\mu
\nabla^\nu \Phi (x) =
\int_\Sigma dS\: n^\mu \sqrt{h} \: \nabla ^\nu 
[A_\nu(x) \lrup{\nabla}_\mu \Phi (x)] \nn\:,
\eeq
where we used the vanishing covariant divergence of the wavefunction.
We can expand the integrand by adding  and subtracting a convenient term,
 obtaining: 
\beq
\int_\Sigma dS\: n^\mu \sqrt{h} \: A_\nu(x) \lrup{\nabla}_\mu
\nabla^\nu \Phi (x) = \nn
\eeq
\beq
=\int_\Sigma dS\: n_\mu \sqrt{h} \:\nabla_\nu
[ A^\nu \nabla^\mu \Phi -  A^\mu \nabla^\nu \Phi]+
\int_\Sigma dS\: n_\mu \sqrt{h} \:\nabla_\nu
[ A^\mu \nabla^\nu \Phi -  \Phi \nabla^\mu  A^\mu] = \nn
\eeq
\beq
=\int_\Sigma dS\: n_\mu \sqrt{h} \:\nabla_\nu G^{\nu\mu}
+ \int_\Sigma dS\: n_\mu \sqrt{h} \: F^{\mu\nu} \nabla_\nu \Phi]\:.\nn
\eeq
We defined $G^{\mu\nu} := A^\nu \nabla^\mu \Phi -  A^\mu \nabla^\nu \Phi$,
$F^{\mu\nu}:= \nabla^\nu A^\mu -\nabla^\mu A^\nu$
and used $\nabla_\nu\nabla^\nu \Phi =0 $
as well as $\nabla_\nu \nabla^\mu A^\nu= \nabla^\mu \nabla_\nu A^\nu=0$
due to the flatness of the space. Notice that, due to Klein-Gordon equations,
$\nabla_\nu F^{\mu\nu}=0$. Thus,
 integrating by parts in the latter integral and re-introducing the
second arguments $x'$ with its covariant derivative, we can write:
\beq
\int_\Sigma dS\: n^\mu \sqrt{h} \: A_\nu(x) \lrup{\nabla}_\mu
\delta W_{\beta}^{\nu\nu'}(x,x') = \nabla^{\nu'}
 \int_\Sigma dS\: n_\mu \sqrt{h}
 \:\nabla_\nu \left( G^{\nu\mu}(x,x') - F^{\mu\nu}(x)\Phi(x,x')\right)\:.\nn
\eeq
The integral in the right hand side, due to the antisymmetry of the integrand
tensor, reduces to 
\beq
\int_\Sigma dx^1dx^2dx^3\: \sum_{i=1,2,3}\partial_i \left[\sqrt{-g}
\left( G^{i0}- \Phi F^{0i}\right)\right]\:. 
\nn \eeq
This vanishes due to the compactness of the spatial support of $A$.

We may conclude  the static  term 
$\delta W^\pm_{\be \:\mu\mu'}(x,x') $
represents  a remaining static {\em gauge ambiguity} which 
does not affect the physical part
of the theory. We can omit this term in $ W^{\pm}_{\be\:\tau\tau'}(x,x')$
and $W^{\pm}_{\be\:\rho\rho'}(x,x')$ or conversely, we can change
 the value $\be$ appearing in $\delta W^{\pm}_{\be\:\mu\mu'}(x,x')$ 
into  a ``wrong'' variable value 
$\be'\neq \be$ without to affect the physics.
 This determines  an {\em one parameter}
 class of possible thermal (and non thermal in the limit 
$\beta\rightarrow +\infty$) Wightman functions
carrying the same physical content. These changes can be implemented
directly in the thermal master function in Eq.s(\ref{mastert1})
and (\ref{mastert2}) 
or in the thermal Schwinger functions in Eq.(\ref{st1}) and (\ref{st2})
where we have:
\beq
\delta S_{\beta'\:\mu\mu'}(x_E,x_E') = \nabla_\mu \nabla'_{\mu'}
 \Phi(x_E,x_E')
\:\:\:\:
\mbox{where}\:\:\:\: \Phi(x,x') := \frac{1}{4\pi\be'}(ss'+\ln \rho \ln \rho')
\label{last}
\eeq  
All these Euclidean time static 
terms  are solutions of Laplace equation away from the conical
 singularity. Thus, the resulting Schwinger functions remain 
Euclidean Green functions of the Laplacian {\em away from the conical tip}. 
No choice of the value $\be'$, $\be'\rightarrow +\infty$ included,
 produces a vectorial 
Green function on the {\em whole} manifold if the period of the manifold 
$\be \neq 2\pi$. 
This is due to the bad behaviour as $\rho \:\:(\rho')\rightarrow 0$
of the terms $S_{\be \: 01'}$ and $S_{\be \: 10'}$ non depending on 
$\delta S_{\be' \mu\mu'}(x,x')$.\\
When the period of the manifold $\be$ takes the value $2\pi$, no conical
singularity appears and this selects just one Schwinger function.
This Schwinger function is the only Green function of the Laplacian
in the class previously considered
defined in the {\em whole} Euclidean manifold.
 This corresponds to the Schwinger function 
 with $\be' \rightarrow +\infty $, i.e., dropping $+1/4\pi\be$ 
in Eq.(\ref{st1}) and the corresponding added 
static term in $S_{\be\:11'}(x,x')$.

\subsection*{D. COINCIDENCE OF QUANTUM PHOTON VACUA}

In the case of a scalar field, the Wightman functions of Minkowski vacuum 
restricted inside of a Rindler wedges coincide with the thermal Wightman
functions with $\be=2\pi$ calculated with respect to the Fulling vacuum.
This is the content of the Bisognano-Wichmann theorem in terms of
Wightman functions \cite{sewell} \cite{haaglibro}.
 This property can be extended on the quantum state
by GNS theorem and similar.  This property also  holds for spin  $1/2$
 in terms of 
Wightman functions at least 
(see for example  Ref.s  \cite{mova,linet}). 
In the case of  photons, despite of the
 gauge ambiguity in defining Rindler Green
functions we have found, 
the  coincidence of Wightman functions 
holds dealing with test
wavefunction corresponding to physical photons
 and also for  photons carrying modes $A^{(G)}$.
Thus,  in the case of photons belonging to the Lorentz
space ${\cal H}_L$.
Notice also that the positivity of the Wightman functions, working with
physical (Lorentz) states, results to be trivially proved due the positivity
of the Minkowski Wightman functions in the case $\be= 2\pi$.\\ 
Following an algebraic approach, one can try to build up a minimal
$*-algebra$ generated through the field operators when they act on
physical wavefunctions and/or Lorentz wavefunctions. 
In this background,  one should try to implement a GNS reconstruction
to extend to the ``physical part''
of the quantum states the local coincidence of
Wightman functions. However, we do not consider these topics in this 
paper.\\ 
One can verify the coincidence of the above considered Wightman functions
inside of the open Rindler wedge
 employing the following way. First one considers the thermal 
Wightman functions defined in Eq.s (\ref{wt1}), (\ref{wt2}), (\ref{wt3}),
 (\ref{wt4})
and (\ref{wt5}), {\em dropping} all the static terms
$\delta W^{\pm}_{\be \mu\mu'}(x,x')$, namely, by considering the limit as
$\be'\rightarrow +\infty$ in the one-parameter
  Wightman functions class previously discussed.
This omission 
does not affect the final result by dealing with 
 test wavefunctions corresponding to states belonging to ${\cal H}_L$.
 Then, one has
to  
translate the obtained functions in Minkowski coordinates. 
 The resulting functions represent
 just the (non thermal) Minkowski Wightman
functions in  Feynman's gauge:
\beq
W^{\pm\mu\mu'}(x,x') = 
\frac{1}{4\pi^2}\frac{\eta^{\mu\mu'}}{|\vec{x}-\vec{x'}|^2-
(t-t'\mp i\ep)^2}\:.
\nn \eeq 
We report just a technical comment. In order to prove the considered identity
using three smeared distributions, it is convenient to work on the Rindler 
Cauchy surface at $\tau (=\tau') = t (=t') = 0$.
 This is a part of a Minkowski Cauchy surface.
Then, one has to prove the coincidence of the Wightman functions
dealing with wavefunctions with a spatial compact support
in $W_R$ (hence, non containing points with $\rho,\rho'=0$)
employing the usual indefinite scalar product.
The result follows noticing that, on the considered Cauchy surface
$\partial_t= \rho^{-1}\partial_\tau$, and the following {\em three smeared 
distributional}
identities hold there (i.e., at $\tau=\tau'=t=t'=0$):
\beq
\frac{1}{2\rho\rho' [\cosh \alpha -\cosh (\tau-\tau'\mp i\ep')] }
 = \frac{1}{|\vec{x}-\vec{x}'|^2- (t-t'\mp i\ep)^2}\:,
\nn
\eeq
\beq
\partial_\tau
\left(\frac{1}{2\rho\rho' [\cosh \alpha -\cosh (\tau-\tau'\mp i\ep')]}\right)
 =\partial_t
 \left(\frac{1}{|\vec{x}-\vec{x}'|^2- (t-t'\mp i\ep)^2}\right)
\:,
\nn
\eeq
\beq
\partial_\tau\partial_{\tau'}
\left(\frac{1}{2\rho\rho' [\cosh \alpha -\cosh (\tau-\tau'\mp i\ep')]}\right)
 =\partial_t\partial_{t'}
 \left(\frac{1}{|\vec{x}-\vec{x}'|^2- (t-t'\mp i\ep)^2}\right)
\:.
\nn
\eeq
Similar reults arise also dealing with Schwinger functions. However,
in that case
an important geometrical difference arises. The Euclidean Rindler coordinates, 
 as the Euclidean
Minkowski coordinates,
cover
the {\em whole} Euclidean section of Minkowski spacetime. Thus, we expect to
find a coincidence of Rindler Schwinger functions and Minkowski
Schwinger functions everywhere.\\
The transformation law from Euclidean Rindler coordinates $(s,\rho, y,z)$
to Euclidean rectangular coordinates $(r^1,r^2,r^3,r^4)$
reads:
\beq
 r^1=\rho\cos s \:,\:\:\:\: r^4 = 
\rho\sin s \:\:\:\:\mbox{and} \:\:\:\: r^2=y,\:\:\:\: r^3=z\:, \nn
\eeq
 In the present  case $\be=2\pi$, the Rindler Schwinger function
of Eq.s (\ref{st1}),(\ref{st2}),(\ref{st3}),(\ref{st4}) and (\ref{st5}), 
more generally containing $\be'$,
 $0<\be'<+\infty$, in the static  term of Eq.(\ref{last}),
are Green functions of the Laplace operator in the 
manifold $\R^4 - \{ (0,r^2,r^3,0) \: \: |\:\: r^2\:, r^3 \in \R  \}$
endowed with the usual flat Euclidean metric.
 If $\be= 2\pi$ no conical singularity appears
and thus no problem arises in defining Laplacian Green functions on the
whole Euclidean manifold.  One may build up
 the  only Green function
defined on the {\em whole} $\R^4$ which decays:
\beq
  S(x_E,x_E')^{\alpha\alpha'} := 
\frac{1}{4\pi^2} \frac{\delta^{\alpha\alpha'}}{\delta_{\mu\nu}
(r^\mu-r'^\mu)(r^\nu-r'^\nu) } \:, \nn
\eeq
This everywhere defined Green function coincide both with
 the only photon Minkowski Schwinger function in the Feynman gauge
which decays as $|r^4|\rightarrow +\infty$ and 
 the Rindler Schwinger function  containing
no Rindler static terms pointed out in the previous section.


\section*{V. SUMMARY}

In this paper we proved that it is possible to build up  
a mathematically  consistent 
canonical theory for a quasi-free photon field propagating 
in the Rindler wedge, based on
a generalization of the Gupta-Bleuler formalism in the Rindler wedge
and  also considering thermal photons. 
We employed a three-smeared formalism, however
 generalizations to a four-smeared
formalism should be straightforward.  
 We proved that the Fulling-Ruijsenaars formalism based
on a (thermal) 
master function can be extended to include the vectorial photon field
recovering properties similar to  those in the  massless scalar case.\\
We proved also that
 the gauge invariance needs more care than in the Minkowskian case,
in particular dealing with the thermal case  (KMS conditions) and studying
the generalization of the Bisognano-Wichmann theorem for photons
in terms of Wightman functions. In fact,  a Rindler non static
gauge ambiguity  coupled with the presence of the conical
singularity appears when $\be \neq 2\pi$. Such a gauge ambiguity 
is not removed also imposing the validity of the Ward identity which arises 
from BRST invariance.\\ In the case $\be=2\pi$,
 we saw that the theory produces the expected  coincidence of
 of the thermal  Rindler Wightman (Schwinger) functions with the
 Minkowski vacuum 
Wightman (Schwinger) functions as far as the ``physical part''
 of those function
is concerned.

\section*{ACKNOWLEDGMENTS}

I would like to thank  especially  Luciano Vanzo,
and also Guido Cognola, Marco Toller and Sergio Zerbini 
for several discussions on some topics related to this paper.

\section*{APPENDIX A}

In this appendix we shall find the normalization coefficients 
$C^{(\alpha,\omega, k_t)}$
of the modes in Eq.s (\ref{buno}), (\ref{bdue}), (\ref{btre}), 
(\ref{bquattro}) using the scalar product $(\:,\:)$ defined in 
Eq.(\ref{prodotto2}).
It can be simply proved  that: 
\beq
(A^{3},A^{4})=0\:.\nn
\eeq
The remaining scalar products of different modes vanish as specifyed
in {\bf section III B}. 
Thus the modes appearing in Eq.s (\ref{buno}), (\ref{bdue}), (\ref{btre}),
(\ref{bquattro}) define
a set of {\em normal to each other} modes. 
Let us normalize them as required by Eq.s (\ref{uno}), (\ref{due}),
(\ref{tre}).\\
The normalized $A_{\mu}^{(2,\omega,k_{t})}$ reads:
\beq
A_{\mu}^{(2,\omega,k_{t})} = \frac{\sqrt{\sinh \pi \omega}}{2 \pi^{2}
k_{\perp}}
(\rho\partial_{\rho}\phi,-i\frac{\omega}{\rho}\phi,0,0) \label{cdue}\:.
\eeq
Let us prove this.
Employing the definition of $(\:,\:)$ we obtain:
\beq
(A^{(2,\omega,k_{t})}, A^{(2,\omega',k'_{t})}) =\nn
\eeq
\beq
=i \int dx_t \frac{d\rho}{\rho } \: C^{\ast (2,\omega,k_t)}C^{
(2,\omega',k'_t)} \{ \frac{1}{\rho} \partial_\tau\phi^{\ast(\omega,k_t)}
(\partial_\rho \rho \partial_\rho \phi^{(\omega',k'_t)}- \frac{1}{\rho}
\partial^2_\tau \phi^{(\omega',k'_t)} ) +\nn
\eeq
\beq
=-\frac{1}{\rho} \partial_\tau\phi^{(\omega',k'_t)}
(\partial_\rho \rho \partial_\rho \phi^{\ast (\omega,k_t)}- \frac{1}{\rho}
\partial^2_\tau \phi^{\ast (\omega,k_t)}) \} =\nn
\eeq
\beq
=-i \int dx_t \frac{d\rho}{\rho } \: C^{\ast (2,\omega,k_t)}C^{
(2,\omega',k'_t)}(\partial_\tau \phi^{\ast (\omega,k_t)} \nabla^2_t 
\phi^{(\omega',k'_t)} -\partial_\tau \phi^{(\omega',k'_t)} \nabla^2_t
\phi^{(\ast, \omega,k_t)} ) =\nn
\eeq
\beq
= \int  \frac{d\rho}{\rho } \: C^{\ast (2,\omega,k_t)}C^{
(2,\omega',k_t)}\: (2\pi)^2 \delta(k_t-k'_t) \:(\omega + \omega')
k_\perp^2 K_{i\omega}(k_t\rho) K_{i\omega'}(k_t\rho)
=\nn
\eeq
\beq
= C^{\ast (2,\omega,k_t)}C^{(2,\omega',k_t)}
\: (2\pi)^2 \delta(k_t-k'_t) \:(\omega + \omega')
k_\perp^2 
\int_0^{+\infty}  \frac{d\rho}{\rho }
K_{i\omega}(k_t\rho) K_{i\omega'}(k_t\rho)\:.\nn
\eeq
Reminding the relation: 
\beq
\int_{0}^{+\infty}
d\rho\: \rho^{-1}\: K_{i\omega}(k_{\perp}\rho) K_{i\omega'}(k_{\perp}\rho)
=\frac{\pi^{2}}{2\omega \sinh \pi\omega}\:\delta(\omega-\omega') 
\label{relazione1}\:,
\eeq  
and choosing the coefficient $C^2$ as real, we find the form (\ref{cdue})
of the mode $A^2$ producing the required ``delta'' normalization.\\ 
In the case of $C^{1}$ we find:
\beq
\frac{1}{C^{\ast (1,\omega,k_{t})}C^{(1,\omega',k'_{t})}}
(A^{(1,\omega,k_{t})},A^{(1,\omega',k'_{t})})=
i \int dx_{t} \frac{d\rho}{\rho} \sum_{a = y,z} A_{a}^{\ast(1,\omega,k_{t})}
\lrup{\partial}_{\tau}A_{a}^{(1,\omega',k'_{t})}\nn \:,
\eeq
where we used also 
the fact that the only non vanishing {\em Christoffel symbols}
which appear in our coordinates 
are $\Gamma^{\tau}_{\tau\rho} = \Gamma^{\tau}_{\rho\tau}=1/\rho$
and $\Gamma^{\rho}_{\tau\tau}= \rho$ ($\tau$ is the Rindler 
time and
$\rho$ is the non trivial space-like  Rindler coordinate: they are  not
generic
indexes).\\ 
Reminding the form of $A^{(\beta,\omega,k_{t})}$ as function of
$\phi$ given in Eq.(\ref{phi}), we obtain:
\beq
\frac{1}{C^{\ast (1,\omega,k_{t})}C^{(1,\omega',k'_{t})}}
(A^{(1,\omega,k_{t})},A^{(1,\omega',k'_{t})})=\nn
\eeq
\beq
=(\omega+\omega')(2\pi)^{2} k_{\perp}^{2} e^{i(\omega-\omega')\tau}
\delta(k_{t}-k'_{t})\int \frac{d\rho}{\rho}K_{i\omega}(k_{\perp}\rho)
K_{i\omega'}(k'_{\perp}\rho)\:.
\eeq
Using the relation in Eq.(\ref{relazione1})
and  choosing the simplest fase, it arises:
\beq
C^{(1,\omega,k_{t})} = \frac{\sqrt{\sinh \pi \omega}}{2\pi^{2}k_{\perp}} =
C^{(2,\omega,k_{t})}\:. \nn
\eeq 
Finally, we have, by inserting this result in Eq.(\ref{buno}):
\beq
A_{\mu}^{(1,\omega,k_{t})} = \frac{\sqrt{\sinh \pi \omega}}{2 \pi^{2} 
k_{\perp}}
(0,0,k_{y}\phi,-k_{x}\phi) \nn\:.
\eeq
Employing similar calculations, we obtain also:
\beq
C^{(4,\omega,k_{t})} = \frac{\sqrt{\sinh \pi \omega}}{2\pi^{2}k_{\perp}} =
C^{(2,\omega,k_{t})}\nn 
\eeq
and thus
\beq
A_{\mu}^{(4,\omega,k_{t})} \equiv  \frac{\sqrt{\sinh \pi \omega}}{2\pi^{2}
k_{\perp}}
(0,0,ik_{x}
\phi,ik_{y}\phi) \label{cquattro}\:.
\eeq
Calculations for the case $C^{(3,\omega,k_{t})}$ are more complicated.\\
Let us start noting that, from Eq.s (\ref{btre}),  
(\ref{atre}) and (\ref{aquattro}): 
\beq
A_{\mu}^{(3,\omega,k_{t})} 
= C^{(3,\omega,k_{t})}
\left[ \frac{A_{\mu}^{(G,\omega,k_{t})} }{C^{(G,\omega,k_{t})}}
-i  \frac{A_{\mu}^{(L,\omega,k_{t})} }{C^{(L,\omega,k_{t})}} \right]\:. \nn
\eeq
Note also that $( A^{(G,\omega,k_{t})},A^{(G,\omega',k'_{t})})=0$
because $F_{A^G}^{\mu\nu}=0$ and $\nabla_\mu A^{G\mu}=0$.\\
Then, choosing:
$C^3=C^{\ast 3}=C^G=C^{\ast G}=C^L=C^{\ast L}$ we find:
\beq
A_{\mu}^{(3,\omega,k_{t})}
=\left[ A_{\mu}^{(G,\omega,k_{t})} 
-i A_{\mu}^{(L,\omega,k_{t})}  \right]\:, \nn
\eeq
and thus (omitting obvious indexes):
\beq
(A^{(3,\omega,k_{t})},A^{(3,\omega',k'_{t})})=
-i(A^G,A'^L ) + i(A^L,A'^G) + (A^L,A'^L) = \nn
\eeq
\beq
=i(A^G,A'^L ) + i(A^L,A'^G) + \frac{C^3C'^3}{C^4C'^4}(A^4,A'^4)\:,\nn
\eeq
where, as we found,
$C^{(4,\omega',k'_{t})}= \sqrt{\sinh \pi\omega}/ 2\pi^2 k^2_\perp$.\\
It follows expanding the formula above 
($i=\rho,y,z$ and there is understood 
a summation
over repeated indexes):
\beq
\frac{1}{C^{(3,\omega,k_{t})}C^{(3,\omega',k'_{t})}}
(A^{(3,\omega,k_{t})},A^{(3,\omega',k'_{t})})=
\nn
\eeq
\beq
=\frac{1}{C^{(3,\omega,k_{t})}C^{(3,\omega',k'_{t})}} \int dx_t
\frac{d\rho}{\rho} \:\: (A^{\ast G}_i F'^{L}_{\tau i}-
A^{\ast G}_\tau \nabla_\mu A'^{L\mu})+ \nn
\eeq
\beq
-\frac{1}{C^{(3,\omega,k_{t})}C^{(3,\omega',k'_{t})}} \int dx_t
\frac{d\rho}{\rho} \:\: (A'^{ G}_i F^{\ast L}_{\tau i}-
A'^{G}_\tau \nabla_\mu A^{\ast L\mu})+
\frac{1}{C^{(4,\omega,k_{t})} C^{(4,\omega',k'_{t})}}\:
\delta(k_t-k'_t)\:\delta(\omega-\omega') \:.\nn
\eeq
Executing the integrals and using Eq.(\ref{relazione1}) we obtain the final
result:
\beq
\frac{1}{C^{(3,\omega,k_{t})}C^{(3,\omega',k'_{t})}}
(A^{(3,\omega,k_{t})},A^{(3,\omega',k'_{t})})=\nn
\eeq
\beq
= (-\frac{2}{C^{(3,\omega,k_{t})}C^{(3,\omega',k'_{t})}}
+ \frac{1}{C^{(4,\omega,k_{t})}C^{(4,\omega',k'_{t})}})
\:\delta(k_t-k'_t)\:\delta(\omega-\omega') \:.\nn
\eeq
We shall take $
C^{(3,\omega,k_{t})}= C^{(4,\omega,k_{t})}$
and thus we have the following normalization relation:
\beq
(A^{(3,\omega,k_t)}
,A^{(3,\omega',k'_t)}) = - \delta(\omega-\omega') \:\delta(k_t-k'_t) \:,
\nn\eeq
where
\beq 
A^{(3,\omega,k_{t})} = \frac{\sqrt{\sinh \pi \omega}}{2 \pi^{2} k_{\perp}}
(-i\omega \phi,\partial_{\rho}\phi,0,0) 
\nn\:.
\eeq
We have found the normalization constant of all the modes:
\beq
C^{(1,\omega,k_{t})}= C^{(2,\omega,k_{t})}=
C^{(3,\omega,k_{t})}= C^{(4,\omega,k_{t})}= C^{(L,\omega,k_{t})}
=C^{(G,\omega,k_{t})}=
\frac{\sqrt{\sinh \pi \omega}}{2 \pi^{2} k_{\perp}}\:.
\nn
\eeq

\section*{APPENDIX B}

In this appendix we shall prove
 Eq.(\ref{fine}) and Eq.(\ref{fine2}).\\
Let us start with Eq.(\ref{fine}).
Note that, because of the trivial dependence on $\tau$
and $\tau'$ of the function $W^{+}$, we can redefine $D_{\tau\tau'}$,
{\em when it acts on} $W^+$, as
\beq
D_{\tau\tau'}= \frac{1}{2}(\partial_{\tau}^{2}+\partial_{\tau'}^{2})
+ \rho \partial_{\rho}\rho'\partial_{\rho'}\nn\:.
\eeq
 Working on a solution of the {\em scalar}
K-G equation like $W^+$ which is a scalar K-G solution in both
arguments also in the present case $\ep>0$, it arises:
\beq
\frac{1}{\rho}\partial^{2}_{\tau} = \partial_{\rho} \rho \partial_{\rho}
+\rho \nabla_{t}^{2}\:, \nn
\eeq                 
where we posed $\nabla_{t}^{2}= \sum_{i=y,z}\partial_{i}^{2}$.
Thus we may write down $D_{\tau\tau'}$ as
\beq
D_{\tau\tau'}= \frac{1}{2}[\rho^{2} \nabla_{t}^{2} + \rho'^{2}
\nabla_{t}^{'2} +
(\partial_{\ln \rho}+\partial_{\ln \rho'} )^{2}] = \frac{1}{2}[\rho^{2}
\nabla_{t}^{2} + \rho'^{2} \nabla_{t}^{2} +
4 \partial^{2}_{\ln \rho\rho'}]\:. \label{inizio}
\eeq
In the latter term we considered as independent variables $u:=\ln(\rho\rho')$
and $v:= \ln (\rho \rho'^{-1})$. These variables appear in the expression
defining $\alpha$, Eq.(\ref{alpha}), and $\alpha$ appears in $W^{+}$ as
precised by Eq.(\ref{W}). \\
Let us consider the action on $W^{+}$ of the last term in the
equation  written above.\\
\beq
\partial_{\ln \rho\rho'} W^{+}(\tau-\tau',u,v,x_t)
= - W^{+} + \frac{\partial W^{+}}{\partial\al}
\frac{\partial \al}{\partial \ln \rho\rho'} =\nn
\eeq
\beq
=-W^{+} + \frac{\partial W^{+}}{\partial\al}
\frac{1}{\sinh \alpha}\frac{\partial \cosh \al}{\partial \ln \rho\rho'}=
-W^{+} - \frac{\partial W^{+}}{\partial\al}
\frac{1}{\sinh \alpha}\frac{x_t^{2}}{2\rho\rho'}\:, \label{primopezzo}
\eeq
where we used the formula of simple proof:
\beq
\frac{\partial \cosh \al}{\partial \ln \rho\rho'}
 = -\frac{x_t^{2}}{2\rho\rho'}\:.\nn
\eeq
Notice  that:
\beq
x_t^{2}\frac{\partial W^+}{\partial x_t^2} = \frac{\partial W^+}{\partial \al}
\frac{x^2_t}{\sinh \al} \frac{\partial \cosh \al}{ \partial x_t^2}=
\frac{\partial W^+}{\partial \al}
\frac{1}{\sinh \al} \frac{x_t^2}{2\rho\rho'}\label{iii}\:.
\eeq
Comparing Eq.(\ref{primopezzo}) with Eq.(\ref{iii}) it arises:
\beq
\partial_{\ln \rho\rho'} W^{+}= -W^+
-x_t^{2}\frac{\partial W^+}{\partial x_t^2}
=  -(1+ x_t^{2}\frac{\partial \:\:\:\:}{\partial x_t^2}) W^+
=  -(1+ \frac{1}{2}|x_t|\frac{\partial \:\:\:\:}{\partial |x_t|}) W^+
\label{secondopezzo}\:.
\eeq
Iterating the process by considering that
$\partial_{\ln \rho\rho'}$ and $ |x_t|\partial_{|x_t|}$ {\em commute},
we obtain 
\beq
\partial_{\ln \rho\rho'}^{2} W^{+}= W^+ + |x_t| \partial_{|x_t|}W^+
+ \frac{x_t^2}{4}
\nabla^{2}_{t}W^+\nn\:,
\eeq
where we used the independence of $W^+$
on the angular variable of 2-vector $x_t$.\\
Substituting the this  expression in Eq.(\ref{inizio}), we find just:
Eq.(\ref{fine})
\beq
D_{\tau\tau'} W^+ = \frac{1}{2}(\rho^2 \nabla^{2}_{t}+\rho'^2\nabla^{2}_{t}+
x_{t}^{2}\nabla^{2}_{t})W^+ + 2W^+ + 2|x_t| \partial_{|x_t|} W^+=\nn
\eeq
\beq
= \rho\rho' \cosh \alpha \nabla^{2}_{t} W^+ + 2W^+ + 2|x_t|
\partial_{|x_t|} W^+
= \rho\rho'\: \nabla^{2}_{t} (\cosh \al W^+)\:, \nn
\eeq
where we used the formulas following from Eq.(\ref{alpha}):
\beq
\partial_{|x_t|} \cosh \alpha(\rho,\rho',x_t) = \frac{2|x_t|}{\rho\rho'}
\eeq
and
\beq
\nabla^2_t \cosh \alpha(\rho,\rho',x_t) = \frac{2}{\rho\rho'}\:.
\eeq

In order to prove Eq.(\ref{fine2})
notice that, because of the dependence of $W^+$ on $\tau-\tau'$, the operator
$D_{\tau\rho'}$ acting on $W^+$ can be written down as
\beq
D_{\tau\rho'}= -\frac{1}{\rho'} (\rho \partial_{\rho}+
\rho' \partial_{\rho'}) \partial_\tau = -\frac{2}{\rho'}
\partial_{\ln \rho\rho'} \: \partial_\tau=
-2\rho \partial_{\rho\rho'} \: \partial_\tau\:.
\nn
\eeq
We considered $U:= \rho\rho'$ and $V:= \rho/\rho'$ as independent variables
above.\\
Furthermore, from the definition of $W^+$, Eq.(\ref{W}), we obtain also:
\beq
\partial_\tau W^+ = \frac{-2(\tau-\tau')\:W^+}{(\tau-\tau'-i\ep)^2-\al^2}\:.
\nn\eeq
And thus we have, posing $T:= \tau-\tau'-i\ep$:
\beq
D_{\tau\rho'}W^+= 2\rho \:\:
\partial_{\rho\rho'}\:\left[ \frac{2T\: W^+}{T^2-\al^2}\right]
= -\frac{\rho T}{\pi^2}\:\:\partial_{\rho\rho'}\:\left[
\frac{\al}{\rho\rho'\sinh\al}
\frac{1}{(T^2-\al^2)^2}\right] =\nn
\eeq
\beq
=-\frac{T\rho}{\pi^2}\left\{ \frac{-1}{\rho^2\rho'^2}\frac{\al}{\sinh\al}
\frac{1}{(T^2-\al^2)^2}+\frac{1}{\rho\rho'}\:\partial_{\rho\rho'}\:\left[
\frac{\al}{\sinh\al}\frac{1}{(T^2-\al^2)^2}\right]
\right\}=\nn
\eeq
\beq
=\frac{T\rho}{\pi^2}\left\{ \frac{1}{\rho^2\rho'^2}\frac{\al}{\sinh\al}
\frac{1}{(T^2-\al^2)^2}+ \frac{x_t^2}{\rho^2\rho'^2}\:
\frac{\partial\:\:}{\partial x_t^2}\:\left[\frac{\al}{\sinh\al}
\frac{1}{(T^2-\al^2)^2}\right] \right\}\:, \nn
\eeq
where we used the formula:
\beq
\frac{\partial\:\:}{\partial \rho\rho'}\:
f\left(\frac{x_t^2}{\rho\rho'}\right)
= -\frac{x_t^2}{\rho\rho'}\:\frac{\partial\:\:}{\partial x_t^2}
\:f\left(\frac{x_t^2}{\rho\rho'}\right)\:.\nn
\eeq
We have
\beq
D_{\tau\rho'}W^+= \frac{T\rho}{\pi^2\rho^2\rho'^2}
\left\{ \frac{\al}{\sinh\al}
\frac{1}{(T^2-\al^2)^2}+ x_t^2\:
\frac{\partial\:\:}{\partial x_t^2}\:\left[\frac{\al}{\sinh\al}
\frac{1}{(T^2-\al^2)^2} \right]\right\}=\\ \nn
\eeq
\beq
=\frac{T\rho}{\pi^2\rho^2\rho'^2}\:\frac{\partial\:\:}{\partial x_t^2}\:
\:\left[ x_t^2\:\frac{\al}{(T^2-\al^2)^2\:\sinh\al}\right]\nn \:.
\eeq
Notice that it holds:
\beq
\frac{\al}{ (T^2-\al^2)^2\:\sinh\al} = \rho\rho'
\:\frac{\partial\:\:}{\partial x_t^2}\: \frac{1}{T^2-\al^2}\:.\nn
\eeq
Substituting this in the latter line we have:
\beq
D_{\tau\rho'}W^+= \frac{T}{\pi^2\:\rho'}
\:\frac{\partial\:\:}{\partial x_t^2}\: \left[ x_t^2\:
\frac{\partial\:\:}{\partial x_t^2}\: \frac{1}{T^2-\al^2}\right] =
\frac{T}{4\pi^2\:\rho'\:|x_t|}\:
\frac{\partial\:\:}{\partial |x_t|}\:\left[ |x_t|\:
\frac{\partial\:\:}{\partial |x_t|}\: \frac{1}{T^2-\al^2}\right]=\nn
\eeq
\beq
=\frac{T}{4\pi^2\:\rho'}\:\nabla^2_t\: \frac{1}{T^2-\al^2}=
-\rho\: (\tau-\tau') \:\nabla^2_t \:
\left(\frac{\sinh \al}{\al} W^+\right)\:.\nn
\eeq
Thus Eq.(\ref{fine2}) has been proved.

\section*{APPENDIX C}

We shall introduce the definition of Wightman functions based on a 
{\em four smeared formalism}, (see for example 
\cite{kaywald} for the scalar case on a curved space).
Wightman functions \\
$<F| \hat A_{\mu}(x) \hat A_{\mu'}(x')|F>$
 are defined within this formalism by imposing:
\beq
<F|\hat{A}(F)\hat A (F')|F> =\nn
\eeq
\beq
=\int_{W_R} d^4x \sqrt{-g(x)} \int_{W_R} d^4x'\sqrt{-g(x')}\:
F^\mu(x) \:F'^{\mu'}(x') \: <F| \hat A_{\mu}(x) \hat A_{\mu'}(x')|F>
\:, \nn
\eeq
where $F_\nu(y)\in C^{\infty}_{0}(W_R)$ for $\nu=0,1,2,3$ and we defined:
\beq
 <F|\hat{A}(F)\hat A (F')|F> := <F| (A_F ,\hat{A}) (A_{F'},\hat A)|F> \:,
\nn\eeq
 The functions $A_F$ are solutions of K-G equation carrying a compact
 support on Cauchy surfaces
 obtained from
 functions $F$  as
\beq
A_F^\mu(x) = \int_{W_R} d^4y\sqrt{-g(y)} \:E(x,y)^{\mu\nu}
\:F_\nu(y)
 \:. \label{AF}
\eeq
$E(x,y)$ is the ``advanced minus retarded'' fundamental solution
of K-G equation (see {\bf section III.B}).
 Formally speaking (see  Ref.s \cite{kaywald,fulling}
for the scalar case):
\beq
E(x,y)_{\mu\nu}:= [\hat{A}_\mu(x),\hat{A}_\nu(y)]\nn\:.
\eeq
Because of the independence
of the quantum state of that function,
we expect to find, employing test functions with support inside of the
open set $W_R$:
\beq
 E(x,y) = E(x,y)_M \label{eem}\:,
\eeq
 The latter two-point function being the Minkowski
advanced minus retarded fundamental solution. 
We have proved this statement in {\bf section III.B}.\\
Notice that $E(x,y)_M$ is
(distributionally) vanishing outside of the light cone at $y$ and this assures
the compactness of the spatial support of the functions $A_F$ whenever
the functions $F$ belong to $C^{\infty}_{0}$. \\
Another important property which can be simply proved  employing Minkowskian 
coordinate through Eq.s (\ref{AF}) and (\ref{eem}) is: 
\beq
\nabla_{\mu} A_F^\mu(x) = \int_{W_R} d^4y\sqrt{-g(y)} \:E(x,y)_S
\:\nabla_\nu F^\nu(y)
 \:, \nn
\eeq
where $E(x,y)_S$ is the {\em scalar} advanced minus retarded
 fundamental solution.\\
Finally,
 notice that $\nabla_\nu F^\nu =0$ implies $\nabla_\mu A_F^\mu=0$.

\bigskip
\bigskip

\begin{description}

\item[${\bf \dagger}$]
E-mail address: moretti@science.unitn.it

\item[{\bf $^{a)}$}] We are employing the
signature $(-1,1,1,1)$ and thus some
sign results to be changed
with respect to Ref.\cite{higuchi} where they used the opposite signature.

\item[{\bf $^{b)}$}]
Remind also the identity $\Gamma^{\mu}_{\mu\nu}=\partial_{\nu}
\{\ln \sqrt{-g}\}$.

\item[{\bf $^{c)}$}]
This is {\em not} the metric in  Rindler
coordinates which is
$g_{\mu\nu} \equiv \mbox{diag}\:(-\rho^{2},1,1,1,)$.

\item[{\bf $^{d)}$}]
As precised above, the stronger assumption of edge-vanishing
test wavefunctions
permits to drop
boudary terms arising by changing the scalar product $(\:,\:)$ with $(\:,\:)'$.
Thus, in the weak sense, one can drop similar boudary terms
also in
Eq.s (\ref{uno}), (\ref{due}), (\ref{tre}) and re-write these in therms of
$(\:,\:)'$.

\item[{\bf $^{e)}$}]
We shall indicate by $\otimes_T$
the topological (Hilbertian) tensorial product.

\item[{\bf $^{f)}$}]
The energy is negative being
$<F|\hat a_0 \hat H\hat a_0^\dagger |F> < 0$. However, the
Hamiltonian
{\em eigenvalues} of the quanta generated by $a^\dagger_0$ are
positive.

\item[{\bf $^{g)}$}]
Remind
that the spatial surfaces
do not contain the points with $\rho=0$ because $W_R$ is an open set.
 The spatial support of
 the considered solution can contain points with $\rho=0$ (the horizons)
only as $|\tau| \rightarrow +\infty $.

\item[{\bf $^{h)}$}]
The $d\rho$ integration, due to
the factor $K_{i\omega}(k_\perp \rho)$, produces a logarithmicallly divergent
function as $k_t\rightarrow 0$ which does not affect this result.

\item[{\bf $^{i)}$}]
These formulae hold
 on the linear manifold
${\cal D}$, dense
in the considered topology,
 containing all the Fock states carrying whatever finite
number of particles. ${\cal D}$ results to be  invariant
under the action of
$\hat a$, ${\hat a}^\dagger$ as well as $M$.

\item[{\bf $^{j)}$}]
Due to $\overline{{\cal M}} = {\cal H}$,
we can approximate all the scalar
product of the states in ${\cal H}$ by complex linear combinations of
Wightman functions
 with $A,A'\in {\cal S}$.

\item[{\bf $^{k)}$}]
See for example Ref.\cite{vladimirov} in part II section 6.5,
 changing the hypotheses of the example h) and using the same proof.

\item[{\bf $^{l)}$}]
If $f( x_t) = \nabla_t^2 g(| x_t|)$, it is sufficient that
  $\frac{\partial g(x_t) }{\partial \ln |x_t| }\rightarrow 0$ as
 $|x_t|\rightarrow +\infty$. This holds in both the examinated cases below.

\item[{\bf $^{m)}$}]
In order to use the following formula it is sufficient,
if $g(x_t)=g(|x_t|)$,
that $g(x_t) \rightarrow 0$ and
$\ln|x_t|\frac{\partial g(x_t)}{\partial \ln|x_t|}
\rightarrow 0$ as $|x_t|\rightarrow \infty$. This holds
in the present case (as well as in the next one) where we have
(as $|x_t|\rightarrow +\infty$)
$g (|x_t|) = \cosh \alpha W^+ \sim
 (\ln|x_t|)^{-1}$
(and
$g(|x_t|)=  \frac{\sinh \alpha}{\alpha} W^+ \sim
 (\ln|x_t|)^{-2}$).

\item[{\bf $^{n)}$}]
In particular, notice that
$V_{\mu\mu'}(x_E,x_E')|_{x_E = x_E'}= g^E_{\mu\mu'}(x_E)$.

\item[{\bf $^{o)}$}]
Use
standard identities as
$\frac{1}{x\pm i\ep} = \mbox{PV}\: \frac{1}{x}\mp i\pi\:\delta(x)$

\item[{\bf $^{p)}$}]
This follows trivially from the Green functions
 calculated in
 Ref.s \cite{birrelldavies,itzykson}.

\item[{\bf $^{q)}$}]
We use in particular the identity following directly
from Eq.s (\ref{runo}).
$2\rho\rho' \:(\cosh(\tau-\tau') - \cosh \alpha) = |\vec{x}-\vec{x}'|^2 -
(t-t')^2 $
and
$\mbox{sign}(t-t') = \mbox{sign}(\tau-\tau')$
holding
for test functions of the variable $x$
with support inside of the closed light cone
at $x'$.

\item[{\bf $^{r)}$}]
Notice that  the
scalar propagator $G_F$ coincides with
the ghost propagator.

\item[{\bf $^{s)}$}]
The  ``local'' temperature $T$ measured by an observer
situated at a fixed spatial point is related to $T_0$ by the Tolman relation
$T=T_0/\sqrt{-g_{00}}$ see Ref. \cite{landaulifsits} .

\item[{\bf $^{t)}$}]
The condition
$\cosh\left(\frac{2\pi}{\be}\al\right)-
\cosh\left(\frac{2\pi}{\be}(\tau-\tau')\right)= 0$ is equivalent to
$\cosh\al-
\cosh(\tau-\tau')=0$ or, employing Minkowskian coordinates,
$|\vec{x}-\vec{x}'|^2-(t-t')^2=0$

\item[{\bf $^{u)}$}]
Notice  that we must define the smooth test functions
requiring also $f(s,\rho,x_t)|_{\rho=0}= f(s',\rho,x_t)|_{\rho=0}$
 whatever $s,s'\in
(0,\be]$ and $x_t \in \R^2$.
 In order to prove the following formula note also ($x_t\in \R^2$)
$ \frac{\rho^2}{(\rho^2+|x_t-x_t'|^2)^2}\rightarrow \pi \delta(x_t-x_t')$ as
$\rho\rightarrow 0^+$.

\end{description}

\end{document}